\documentclass[12pt]{article}
\pdfoutput=1 

\usepackage{jheppub} 
\usepackage{graphicx}  
\usepackage{dcolumn}   
\usepackage{bm}        
\usepackage{amssymb}   
\usepackage{subfigure}
\usepackage{epstopdf}
\usepackage{comment}
\usepackage{color}
\usepackage{slashed}
\usepackage[utf8]{inputenc}
\usepackage{multirow}
\usepackage{footnote}
\usepackage{amsmath}
\allowdisplaybreaks[4]
\usepackage{accents}
\newlength{\dhatheight}

\usepackage{hyperref}

\def\figureautorefname~#1\null{Fig.\,#1\null}
\def\tableautorefname~#1\null{Tab.\,#1\null}

\def\equationautorefname~#1\null{Eq.\,(#1)\null}

\title{Search for the Electric Dipole Moment and anomalous magnetic moment of the tau lepton at tau factories}
\author[a,b,c]{ Xin Chen }
\author[d,a]{ Yongcheng Wu }
\affiliation[a]{Department of Physics, Tsinghua University, Beijing 100084, China}
\affiliation[b]{Collaborative Innovation Center of Quantum Matter, Beijing 100084, China}
\affiliation[c]{Center for High Energy Physics, Peking University, Beijing 100084, China}
\affiliation[d]{Ottawa-Carleton Institute for Physics, Carleton University, \\
1125 Colonel By Drive, Ottawa, Ontario K1S 5B6, Canada}
\emailAdd{xin.chen@cern.ch}
\emailAdd{ycwu@physics.carleton.ca}

\abstract{
Precise measurement of the Electric Dipole Moment (EDM) and anomalous magnetic moment ($g$-2) of particles are important tests of Beyond Standard Model (BSM) physics. It is generally believed that the tau lepton couples more strongly to BSM due to its large mass, and can be searched for at collider experiments. A new method to approximately reconstruct the neutrinos from the hadronic decays of $\tau^-\tau^+$ pairs produced at $e^-e^+$ tau factories is proposed. With all final state particle momenta available, observables based on matrix elements and sensitive to BSM are calculated. It is estimated that with 50 ab$^{-1}$ of data to be delivered by the $Belle$-II experiment, a tau EDM search with a 1-$\sigma$ level precision of $|d_\tau^{NP}|<2.04\times 10^{-19}$ e$\cdot$cm, and $g$-2 search with $|a_\tau^{NP}|<1.75\times 10^{-5}$ ($1.5\%$ of the SM prediction), can be expected when systematics are not considered. The new precision can effectively constrain BSM models with heavy mirror neutrinos. It can also constrain models containing a light scalar with mass at $O$(1 GeV), which can explain the current muon $g$-2 anomaly as well. The method in this work offers a new opportunity to search for BSM at current and future tau factories with high precision.
}

\begin{document}
\titlepage
\maketitle
\newpage

\flushbottom

\section{Introduction}

When BSM exists in the loop diagrams of the photon-lepton interaction vertex, e.g. in some sypersymmetric model and extended technicolor model etc.~\cite{Zhao:2014vga,Yamanaka:2014nba,Appelquist:2004mn,Ibrahim:2008gg,Ibrahim:2010va,Ilakovac:2013wfa}, the lepton can possess extra EDM ($d_\ell$) and/or anomalous magnetic moment ($a_\ell = (g-2)/2$). The most general vertex function describing relevant interactions between lepton and photon can be written in the form~\cite{Eidelman}:
\begin{align}
\label{equ:GammaLepton}
\Gamma^\mu(q^2) = -ieQ_\ell\left\{\gamma^\mu F_1(q^2) + \frac{\sigma^{\mu\nu}q_\nu}{2m_\ell}\left[iF_2(q^2)+F_3(q^2)\gamma_5\right]+\left(\gamma^\mu-\frac{2q^\mu m_\ell}{q^2}\gamma_5F_4(q^2)\right)\right\}
\end{align}
where $m_\ell$ is the mass of the lepton, $eQ_\ell$ is the corresponding charge, $\sigma_{\mu\nu}=i[\gamma_\mu, \gamma_\nu]/2$, and $q$ is the ingoing four-momentum of the photon. 

In general, the contributions from BSM to these form factors ($F_{1,2,3,4}(q^2)$) can be analyzed in the framework of effective field theory (EFT) where the SM is extended by a set of higher-dimension operators that are suppressed by the new physics scale $\Lambda$. Any deviation from the SM predictions in these form factors which can be linked to physical observables will be a direct hint of BSM. 

In the limit of $q^2\to0$, $F_2(q^2)$ and $F_3(q^2)$ are directly related to the anomalous magnetic moment and EDM:
\begin{align}
\label{equ:EDM_g2_Def}
a_\ell = F_2(0),\quad d_\ell = \frac{eQ_\ell}{2m_\ell}F_3(0)
\end{align}
Strong constraints have been set on the electron EDM by experiments such as ACME~\cite{ACME,Andreev:2018ayy}. The anomalous magnetic moment of the muon has also been measured at Brookhaven~\cite{g_minus_2a,Miller:2007kk} which showed a deviation from the SM value at about 3.5$\sigma$, and it is about to be measured more precisely by the Fermilab $g$-2 experiment~\cite{g_minus_2b,Grange:2015fou,Chapelain:2017syu} and another one under preparation at J-PARC~\cite{Abe:2019thb}. 
Due to the rapid decay of the tau lepton, searching for its BSM signatures is difficult but can still be carried out at collider experiments~\cite{tau_exp1,Abdallah:2003xd,Ananthanarayan:1994af,Ananthanarayan:2002fh,Bernabeu:2004ww,Bernabeu:2006wf,Bernabeu:2007rr,Bernabeu:2008ii,Atag:2010ja,Billur:2013rva,Ozguven:2016rst,Koksal:2017nmy}. The current best measurements, at 95\% confidence level (CL), are~\cite{tau_exp1,Abdallah:2003xd,Patrignani:2016xqp}:
\begin{eqnarray}
-2.2<&\ \Re(d_\tau)\ &<4.5~(10^{-17}\text{e}\cdot\text{cm}), \nonumber \\
 -0.052<&a_\tau&<0.013. 
\end{eqnarray}
Compared with the EDM measurement of the electron ($<1.1\times10^{-29}\text{e}\cdot\text{cm}$), the current precision for tau is ten orders of magnitude lower. The SM prediction of $g$-2 for tau is~\cite{Eidelman:2007sb}:
\begin{equation}
a_\tau = 117721(5)\times10^{-8},
\end{equation}
which is about an order of magnitude below the current experiment precision. Thus, any progress in these measurements will be crucial for the searching of BSM. 

However, in contrast to electron and muon $g$-2 measurements, where the exchanged photon is nearly on-shell ($q^2=0$), in the collider based measurement of $\tau$ lepton, the photon is off-shell. In this case it is the form factors at corresponding scale ($F_2(q^2)$ and $F_3(q^2)$), not $a_\tau$ and $d_\tau$, that are measured~\cite{Bernabeu:2007rr}. However, when the BSM scale $\Lambda^2$ is much higher than $q^2$ of the collision process, terms of higher orders of $q^2/\Lambda^2$ can be neglected in the form factor expansion, and the collider measurement can be interpreted through~\autoref{equ:EDM_g2_Def} into constraints on $d_\tau^{NP} = d_\tau - d_\tau^{SM}$ and $a_\tau^{NP} = a_\tau - a_\tau^{SM}$. When the BSM scale is lower than the measurement energy scale
(as illustrated by the light scalar model in~\autoref{sec:model}), the real part of the contribution to the form factor from loop processes is used to constrain the BSM physics.

It is expected that billions of tau pairs will be produced at the $e^-e^+$ colliders, such as $Belle$-II~\cite{BelleII} and future tau-charm factories. Because the chirality of leptons is preserved in their interactions with photons, the two taus are expected to have spins well correlated with each other. On the other hand, different interactions in~\autoref{equ:GammaLepton} may cause different spin correlations that can be detected in the topology of the tau decay products. The tau EDM and $g$-2 are thus searched for in this work, based on this idea.

This paper is organized as follows. First the method that is used to reconstruct the neutrinos from the $\tau$ decay is introduced in \autoref{sec:reconstruction}, where the simulation of the signal and backgrounds and the event selection are also discussed. Then the matrix element for each event is calculated and the optimal observable is constructed to measure the tau EDM and $g$-2 in \autoref{sec:measurement}. The constraints on two example BSM models are obtained using the tau EDM and $g$-2 measurement in  \autoref{sec:model}. The main results are summarized in \autoref{sec:summary}.

\section{Event selection and reconstruction of neutrinos from $\tau$ decay}
\label{sec:reconstruction}

Tau leptons are quite complicated objects at colliders, mainly because the neutrinos in their decay products are undetectable. While the spin correlation can be partially obtained with only visible decay products~\cite{Comp-Phys-Com-64-275-1991,Tsai:1971vv}, the information of neutrinos is important to help improve the reconstruction of the tau spin correlation. The technique to reconstruct the neutrino momenta from tau decays developed in our previous works~\cite{CEPC_hcp,Chen:2017nxp} is used for the low energy $e^-e^+$ collisions which makes use of the full information per event including the impact parameters of charged tracks from tau decays. 
Because $Belle$-II started taking data with the full detector in 2019~\cite{BelleII_talk}, it is worthwhile to simulate data with the $Belle$-II detector, and investigate the sensitivity one can expect from it for both tau EDM and $g$-2 searches.

\subsection{Simulation of the Signal and Backgrounds}

Although $Belle$-II is a $B\bar{B}$ factory operating at the $\Upsilon(4S)$ resonance energy, it is also a factory for $\tau^-\tau^+$ pairs, with $\sim 10^{10}$ pairs produced each year till 50 ab$^{-1}$ of data is collected by the middle of next decade. This offers an opportunity to search for BSM with high precision that has not been reached so far. The data analyzed in this work is simulated with the asymmetric energy of 7 (4) GeV for the electron (positron) beam as for $Belle$-II. The beams are assumed to be unpolarized.  Both the $\tau^-\tau^+$ signal and $q\bar{q}$ continuum background (with $q=d,u,s,c$) are generated with MadGraph5~\cite{MG5}, parton showered and hadronized by Pythia8~\cite{Pythia8}. To preserve the tau spin correlation, the tau decay models from~\cite{Hagiwara:2012vz} are used to decay the tau lepton inside MadGraph5. The photon ISR/FSR effects are simulated by Phythia as well\footnote{However, the photon emission for tau decay products is not modeled, whose effect is expected to be small since hard photon emission is rare and the analysis vetoes extra photons in the event.}. The $\Upsilon(4S)\to B\bar{B}$ process is another background, which is generated with EvtGen~\cite{EvtGen}. The events are afterwards passed through DELPHES 3.4.0~\cite{Delphes} simulating the detector acceptance and response of $Belle$-II. Tracking and calorimetry are limited in pseudorapidity to roughly $-1.32<\eta<1.90$. A few key detector parameters are worth noting. The track momentum resolution follows $0.3\%\oplus 0.1\%(p_\text{T}/\text{GeV})$~\footnote{Due to the precise track direction measurements, there is no need to smear the track directions. It is nevertheless checked that when track $\eta$ and $\phi$ are smeared with a resolution of 0.001, the impact on neutrino reconstruction is negligible.}, and the impact parameters follow the resolution of $a \oplus b/(p_\text{T}\sin\theta^{1/2})$, with $a$=0.015 (0.020) mm and $b$=0.007 (0.010) mm$\times$GeV for the transverse (longitudinal) impact parameters, $d_0(z_0)$. The particle ID efficiencies are parameterized according to~\cite{BelleII}, but the performance of $BABAR$~\cite{BaBar} is also referred to when these parameters are not available in the former, especially for low $p_\text{T}$ charged tracks for which the Cherenkov detector is not efficient, and $dE/dx$ is then used for these cases. The backward region with $-1.32<\eta<-0.75$ is not covered by the Cherenkov detector, and kaons in this region with $p_\text{T}>$0.7 GeV are by default identified as pions. In most other cases, the kaon and pion ID efficiencies are high and their mix-ID rates are at a few percent level.

The major hadronic tau decay modes considered in this work and their branching fractions are 
\begin{align}
\label{equ:taudecaychannel}
\tau^\pm&\to\pi^\pm\nu\quad(10.8\%), \nonumber \\
\tau^\pm&\to\pi^\pm\pi^0\nu(\rho^\pm)\quad (25.4\%),\nonumber \\
\tau^\pm&\to\pi^\pm\pi^\pm\pi^\mp\nu(a^\pm)\quad (9.3\%),
\end{align}
and the six combinations of these tau decay modes are used in the analysis.

The tracks are required to have a minimum $p_\text{T}$ of 0.2 GeV, and neutral clusters have a minimum energy of 0.1 GeV. To find the two hadronic tau candidates, the charged tracks are divided into two sets, with a net charge of $\pm$1. For three-prong taus and events with neutral pions, the combinations with masses closest to $a_1$ and $\rho$ mesons are chosen, and the events with $a_1$ and $\rho$ candidate mass larger than the nominal tau mass are rejected.

\begin{figure}[!htb]
\centering
\includegraphics[width=0.9\textwidth]{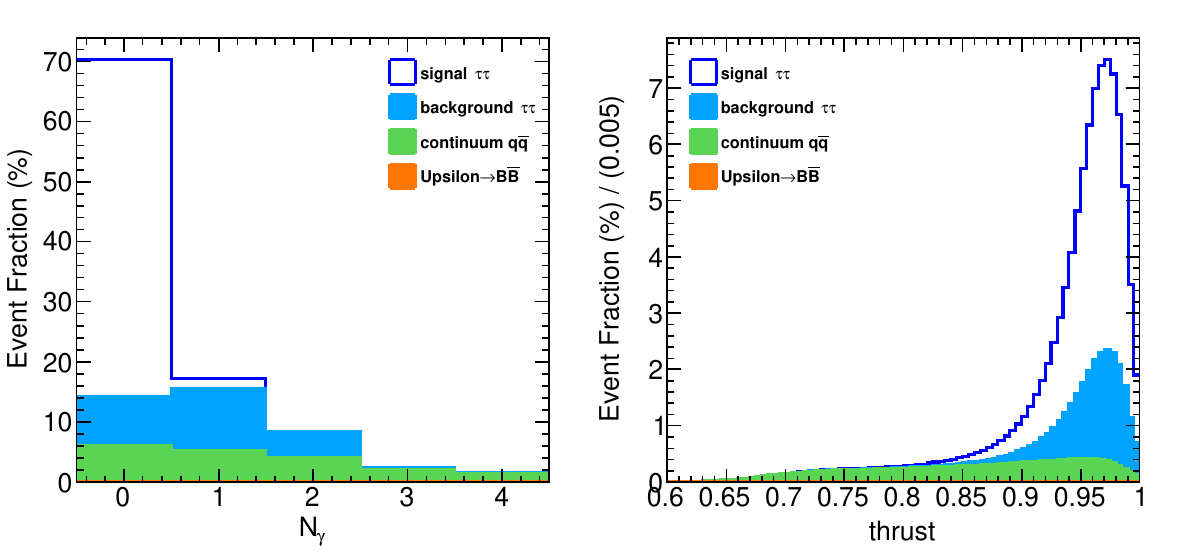}
\caption{The $N_\gamma$ and thrust distributions after the tau pair selection. Different simulation components are stacked. }
\label{fig:ngam_thrust}
\end{figure}

To suppress backgrounds, events with identified $e^\pm$, $\mu^\pm$, $K^\pm$, $p^\pm$ and $K^0_L$ are rejected. The numbers of $\pi^\pm$ and $\pi^0$ for each tau candidate should match one of three tau decay modes in \autoref{equ:taudecaychannel}. Since the impact parameter resolutions degrade fast at low $p_\text{T}$, it is further required that each tau should have at least one track with $p_\text{T}>$0.4 GeV. 
For events with $\pi^0$'s, there will be extra photon candidates due to unidentified $\pi^0$, or one of the photons from its decay is out of detector acceptance. Requiring $N_\gamma=0$ can suppress a lot of the continuum background as well as the background from $\tau^+\tau^-$ production itself. The latter refers to the wrongly reconstructed tau decay modes in the $\tau^-\tau^+$ signal itself, which mainly come from unidentified $\pi^0$'s. The left panel of \autoref{fig:ngam_thrust} shows the $N_\gamma$ distribution before the cut.

\begin{figure}[!tb]
\centering
\includegraphics[width=0.45\textwidth]{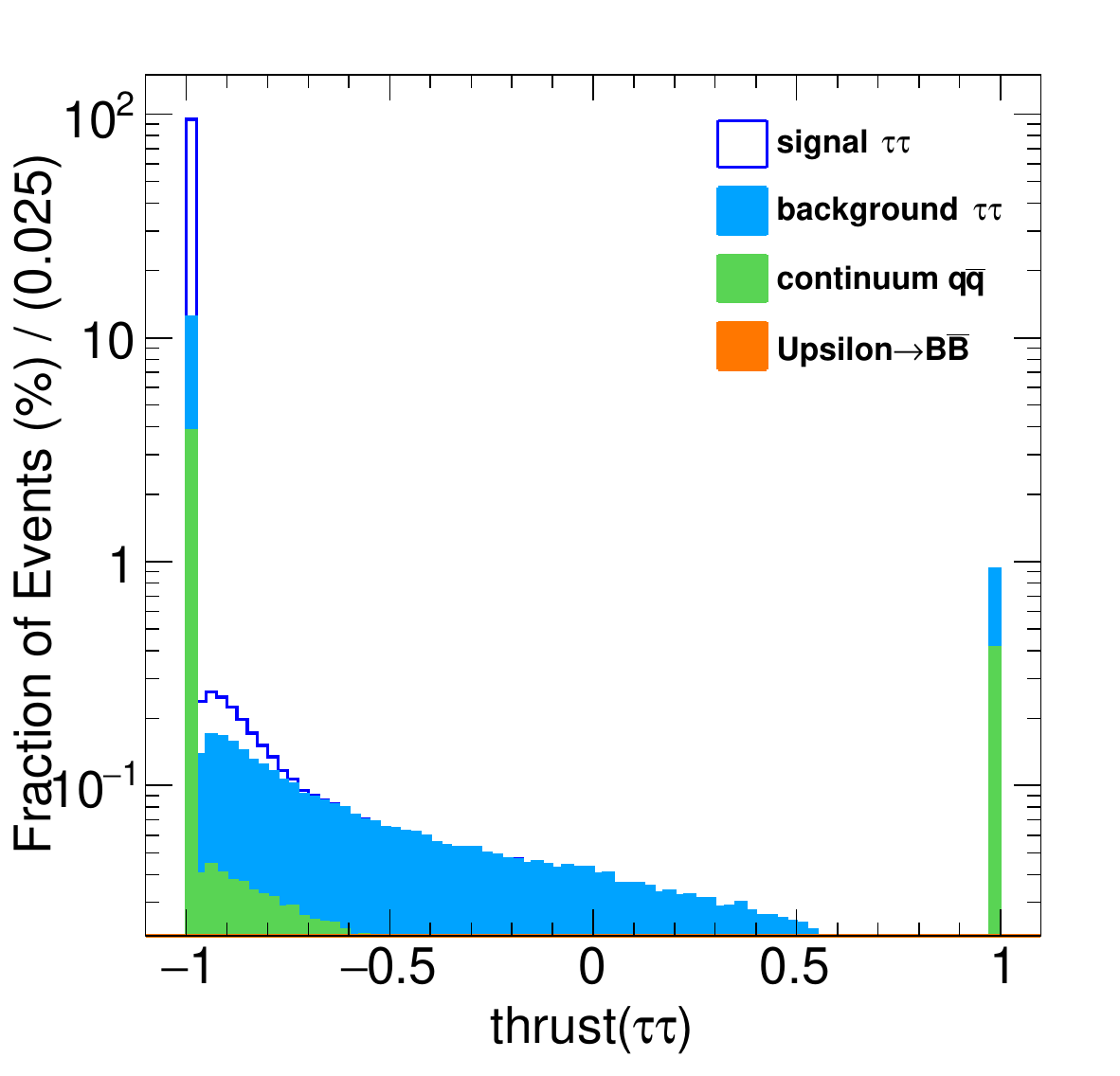}
\caption{The distribution of $\text{thrust}_N(\tau_1)\cdot\text{thrust}_N(\tau_2)$ after the tau pair selection, $N_\gamma$ and thrust cuts. Different simulation components are stacked. }
\label{fig:thrust_normalized}
\end{figure}

It is also required that in the center-of-mass (CMS) frame, the event thrust~\cite{Farhi:1977sg} should satisfy thrust$>$0.85, whose distribution is also shown in the right panel of \autoref{fig:ngam_thrust}. The event thrust is defined as

\begin{equation}
\label{equ:thrust}
\text{thrust} = \text{max}_{\bold{n}_{\text{thr}}} \frac{\sum_{i}\left| \bold{n}_{\text{thr}}\cdot\bold{p}_i\right|}{\sum_{i}\left|\bold{p}_i\right|},
\end{equation}
where $\bold{n}_{\text{thr}}$ is a unit vector named thrust axis, which is defined in the CM frame (a Lorentz boost with $\beta_\text{z}=-3/11$ is applied to all visible tau decay products in order to transform to this frame), $\bold{p}_i$ is the momentum of each visible (charged or neutral) tau decay product in CM frame. The thrust axis is found such that the maximum value of thrust is achieved. To further suppress backgrounds, a normalized thrust is defined for each tau candidate as

\begin{equation}
\label{equ:thrust_normalized}
\text{thrust}_N(\tau_i) = \frac{\sum_{j} \bold{n}_{\text{thr}}\cdot\bold{p}_{ij} }{\sum_{j}\left|\bold{n}_{\text{thr}}\cdot\bold{p}_{ij}^+\right|+\sum_{j}\left|\bold{n}_{\text{thr}}\cdot\bold{p}_{ij}^-\right|},
\end{equation}
where $\bold{p}_{ij}^{+}$ ($\bold{p}_{ij}^{-}$) is the momentum of the $j$'th decay product of the $i$'th tau candidate, whose projection onto thrust axis $\bold{n}_{\text{thr}}$ is positive (negative). The requirement $\text{thrust}_N(\tau_1)\cdot\text{thrust}_N(\tau_2)$=$-1$ can further suppress the continuum background and the $\tau^+\tau$ background by about $29\%$, while signal $\tau^+\tau^-$ is only decreased by $<1\%$. The distribution of this variable is shown in \autoref{fig:thrust_normalized}.

The effective cross sections of different processes after all the cuts are listed in \autoref{tab:events}. The $\Upsilon(4S)$ background is severely suppressed due to $K^\pm$ and $K^0_L$ veto and the thrust cut. After all selection cuts, the total background (including the wrong $\tau^+\tau^-$ modes) constitutes about $12.5\%$ of the total events. Thus one is left with a clean sample of well reconstructed $\tau^-\tau^+$ pairs for the BSM search.

\begin{table}[!hbt]
\centering
\caption{The effective cross sections for different processes after the selection cuts, in different $\tau^+\tau^-$ decay modes.}
\label{tab:events}
\begin{tabular}{ccccc}
\hline
Mode & Signal $\tau^+\tau^-$ (pb) & Background $\tau^+\tau^-$ (pb) & Continuum (pb) & Upsilon (fb) \\
\hline
$a_1 + a_1$ & 3.09 & 0.00 & 0.22 & 0.37 \\
$a_1+\rho_{\phantom{1}}$ & 16.14 & 0.39 & 0.73 & 1.16 \\
$a_1+\pi_{\phantom{1}}$ & 9.30 & 0.70 & 0.42 & 0.59 \\
$\pi_{\phantom{1}}+\pi_{\phantom{1}}$ & 7.42 & 2.50 & 0.51 & 0.68 \\
$\pi_{\phantom{1}}+\rho_{\phantom{1}}$ & 24.13 & 3.16 & 0.98 & 1.01 \\
$\rho_{\phantom{1}}+\rho_{\phantom{1}}$ & 20.96 & 1.20 & 0.73 & 1.19 \\ \hline
Total & 81.04 & 7.95 & 3.58 & 4.99 \\
\hline
\end{tabular}
\end{table}

\subsection{Reconstruction of the neutrinos}

\begin{figure}[!htb]
\centering
\includegraphics[width=0.2\textwidth]{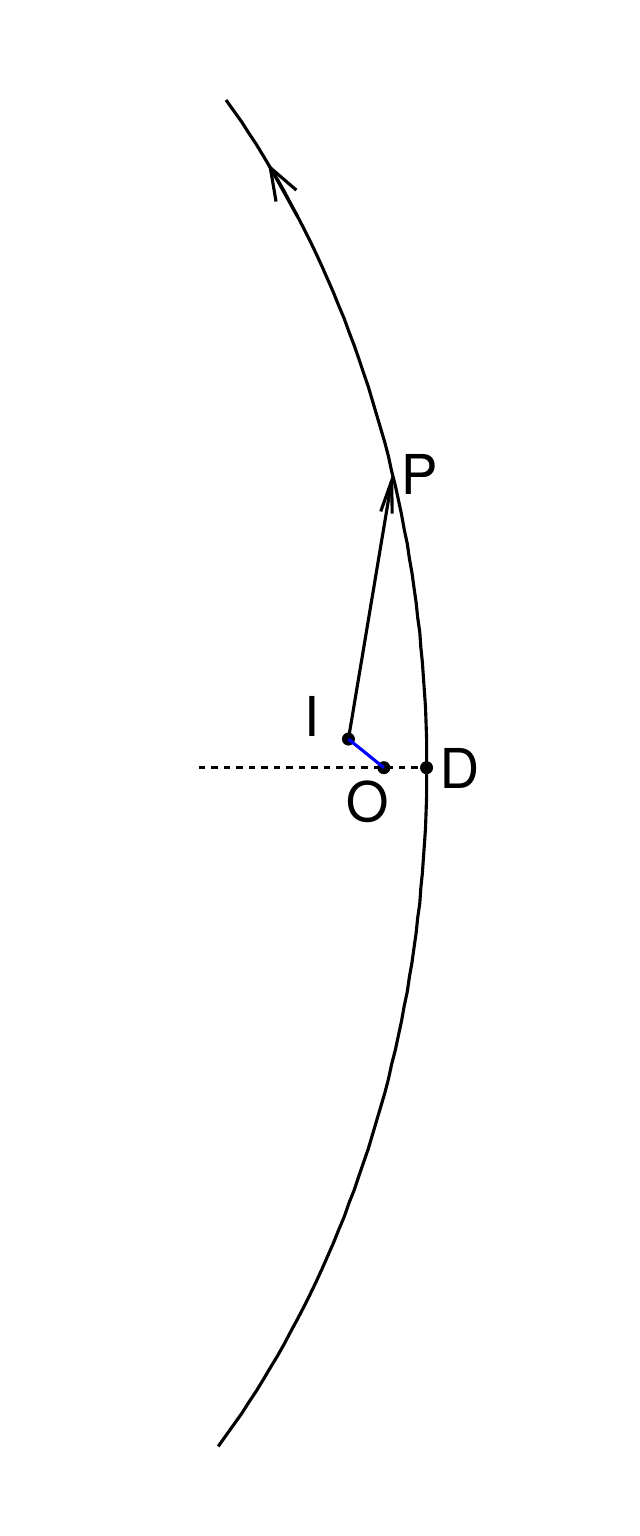}
\put(-30, 40){\textbf{(a)}}\hspace{15mm}
\includegraphics[width=0.2\textwidth]{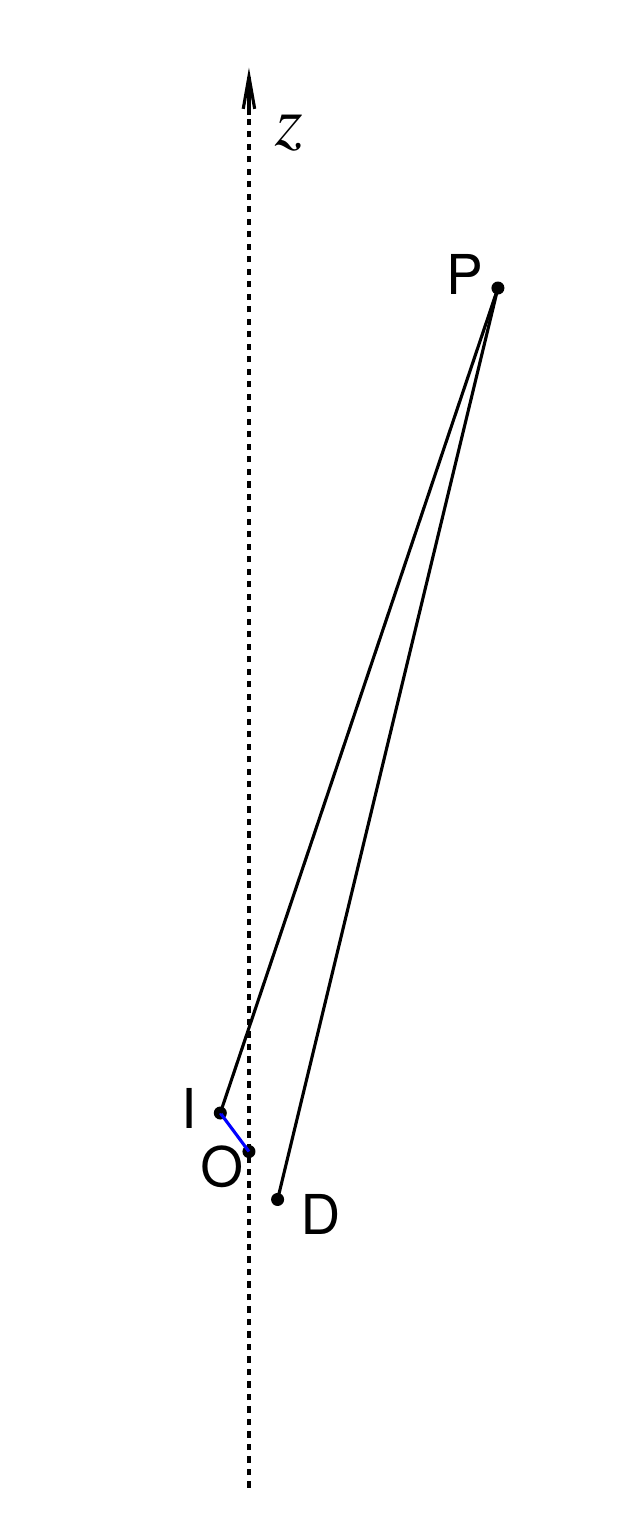}
\put(-30, 40){\textbf{(b)}}
\caption{ Illustration of a track trajectory viewed in the (a) $x$-$y$ and (b) $r$-$z$ planes, where $O$ is the origin, $I$ is the interaction point, $P$ is the tau decay point, and $D$ is the track perigee with respect to $O$.  }
\label{fig:demo_imp}
\end{figure}

In each event, the two missing neutrinos contain six free parameters, while the tau mass\footnote{The tau mass constraints do not appear explicitly in the $\chi^2$, and are only used in the calculation of the tau momentum vectors at the intermediate step.}, the total four-momentum of the event, and the impact parameter measurements can provide at least eight constraints which are sufficient to determine these six free parameters. In this work, a $\chi^2$ fitting is performed event by event, with each constraint contributing one term to the overall $\chi^2$:
\begin{equation}
\label{eq:eq4}
 \chi^2 = \chi^2_\text{evt} + \chi^2_\text{IP} + \chi^2_\text{Imp},
\end{equation}
where
\begin{align}
 \label{eq:chi2_1} 
  \chi^2_\text{evt} & =  \sum_{i=0}^{3}\left(\frac{p_{\text{evt},i}^{\text{fit}}-p_{\text{evt},i}}{\sigma_{\text{evt},i}}\right)^2, \\
  \label{eq:chi2_2} 
  \chi^2_\text{IP} & =  \left(\frac{x_{\text{I}}^{\text{fit}}-x_{c}}{\sigma_x}\right)^2  + \left(\frac{y_{\text{I}}^{\text{fit}}-y_{c}}{\sigma_y}\right)^2  +\left(\frac{z_{\text{I}}^{\text{fit}}-z_{c}}{\sigma_z}\right)^2, \\
  \chi^2_\text{Imp} & =  \sum_i\chi_{\text{Imp},i}^2.
  \label{eq:chi2_3} 
\end{align}
\autoref{eq:chi2_1} sums over the $\tau^-\tau^+$ system four-momentum $p_\text{evt}=(11,0,0,3)$ GeV, and $\sigma_{\text{evt},i}$ are expected resolutions. In~\autoref{eq:chi2_2}, ($x_{\text{I}}^{\text{fit}}$, $y_{\text{I}}^{\text{fit}}$, $z_{\text{I}}^{\text{fit}}$) are the coordinates of the interaction point (IP)\footnote{The IP coordinates here are the coordinates of the event-by-event interaction point.}, and ($x_c$, $y_c$, $z_c$) the coordinates of the beam spot center with $\sigma_{x,y,z}$ being their resolutions. Each track $i$ contributes to \autoref{eq:chi2_3} a term of the form~\footnote{Potential correlations between the impact parameters and track momentum are simulated, and little impact on the neutrino fit (with the same $\chi^2$ set-up) is found. Therefore these correlations are neglected in both the simulation and $\chi^2$.}
\begin{equation}
\chi^2_\text{Imp,i} = \left(\frac{d_{0,i}^{\text{fit}}-d_{0,i}}{\sigma_{d_{0,i}}}\right)^2 + \left(\frac{z_{0,i}^{\prime}-z_{0,i}}{\sigma_{z_{0,i}}}\right)^2,
\label{eq:eq6}
\end{equation}
where impact parameters $d_0$ and $z_0$ are measured with respect to the origin~\footnote{Similar $\chi^2_\text{Imp}$ expressions for tracks whose trajectories do not intersect with the tau flight direction (due to resolution effects) can be found in~\cite{CEPC_hcp,Chen:2017nxp}}.
The variables with a superscript ``fit" denote fitted values by minimizing~\autoref{eq:eq4}. The tau momentum vectors can be fully derived from the $\tau^-\tau^+$ system total four-momentum (subject to resolutions in~\autoref{eq:chi2_1}) and the nominal tau mass, up to a two-fold ambiguity~\cite{Kuhn}. The $z_{0}^{\prime}$ is a function of IP, $d_{0}^{\text{fit}}$ and tau momentum vector. As illustrated in~\autoref{fig:demo_imp}(a), knowing these quantities, the tau decay point can be determined in the transverse plane, and the $z_{0}^{\prime}$ can be then determined by
\begin{equation}
z_0^{\prime} = z_{\text{I}} + L \sinh \eta_\tau - S \sinh \eta_{\text{track}},
\label{eq:eq7}
\end{equation}
where $L$ is transverse distances between $I$ and $P$, and $S$ is the trajectory length from $P$ to $D$ in the transverse plane too. \autoref{fig:demo_imp}(b) illustrates the $z$-positions of the involved points. When $z_{0}^{\prime}$ is compared with the original $z_{0}$ from track fitting, an additional constraint is formed.

The parameters $\sigma_{\text{evt},i}$ are optimized to give the best neutrino reconstruction per $\tau^+\tau^-$ decay mode\footnote{They are tuned such that the fraction of events with good reconstructed neutrinos is the largest, and they also control the relative weight of~\autoref{eq:chi2_1} with respect to \autoref{eq:chi2_2}-\autoref{eq:chi2_3}.}, and the result is given in~\autoref{tab:sigma}. The fact that $\sigma_{\text{evt},E}$ and $\sigma_{\text{evt},z}$ are five times larger than $\sigma_{\text{evt},x}$ and $\sigma_{\text{evt},y}$, is based on the consideration of the ISR effect where either $e^+$ or $e^-$ radiates a hard collinear photon in the initial state. The beam spot center can be measured by consecutive events in a few luminosity blocks with multiple tracks (e.g., from $q\bar{q}$) in which a primary vertex can be well reconstructed. In this work, we assume that the collision always happens at the origin. Therefore, $x_c$ and $y_c$ are smeared according to the transverse beam profile parameters ($\sigma_x=10$ $\mu$m and $\sigma_y=60$ nm from~\cite{BelleII}) around the origin to model the beam spot spread in the plane transverse to the beams. The beam bunch length (6 mm) is too big to constrain $z_{\text{I}}$. Instead, it is found that the average of impact parameters, $z_c = (\sum_{i=1}^{n} z_{0,i})/n$, is a good estimation of $z_c$, where $n$ is the total number of charged tracks in the event. The corresponding resolutions ($\sigma_z$) are given in the last row of \autoref{tab:sigma}.

\begin{table}[!hbt]
\centering
\caption{The optimized resolutions for the resolution parameters that enter \autoref{eq:eq4} in different $\tau^+\tau^-$ decay modes.}
\label{tab:sigma}
\begin{tabular}{c|c|c|c|c|c|c}
\hline
 & $a_1 + a_1$ & $a_1+\rho$ & $a_1+\pi$ & $\pi+\pi$ & $\pi+\rho$ & $\rho+\rho$ \\
\hline
$\sigma_{\text{evt},x}$, $\sigma_{\text{evt},y}$ (MeV) & \multicolumn{3}{c|}{5} & \multicolumn{3}{c}{10} \\ \cline{2-7}
$\sigma_{\text{evt},E}$, $\sigma_{\text{evt},z}$ (MeV) & \multicolumn{3}{c|}{25} & \multicolumn{3}{c}{50} \\ \hline
$\sigma_z$ ($\mu m$) & 23 & 27 & 27 & 40 & 42 & 42 \\
\hline
\end{tabular}
\end{table}



In the per-event minimization of~\autoref{eq:eq4}, the fitted parameters are the system four-momentum $p_{\text{evt},i}^{\text{fit}}$, the impact parameter $d_{0,i}^{\text{fit}}$ of each track ($z_{0,i}^{\prime}$ is a dependent quantity), and the IP coordinates ($x_{\text{I}}^{\text{fit}}$, $y_{\text{I}}^{\text{fit}}$, $z_{\text{I}}^{\text{fit}}$), within their respective uncertainties. 
Their fitted values are first scanned over to find a coarse global minimum, after which MINUIT~\cite{MINUIT} is performed around this point for a better estimation. There are two steps behind minimizing~\autoref{eq:eq4}. First, for a given set of values for $p_{\text{evt},i}^{\text{fit}}$, the four-momentum of each tau can be calculated by transforming the system to the ditau center-of-mass frame, but up to a two-fold ambiguity~\cite{Kuhn}. Second, the taus are boosted back to the lab frame, and \autoref{eq:chi2_2}-\autoref{eq:chi2_3} are evaluated for each of the two solutions. The solution with the smaller combined value of \autoref{eq:chi2_2} and \autoref{eq:chi2_3} is chosen. The neutrino four-momentum is then easily obtained by subtracting the four-momenta of the visible decay products from the tau.

\begin{figure}[!tb]
\centering
\includegraphics[width=0.9\textwidth]{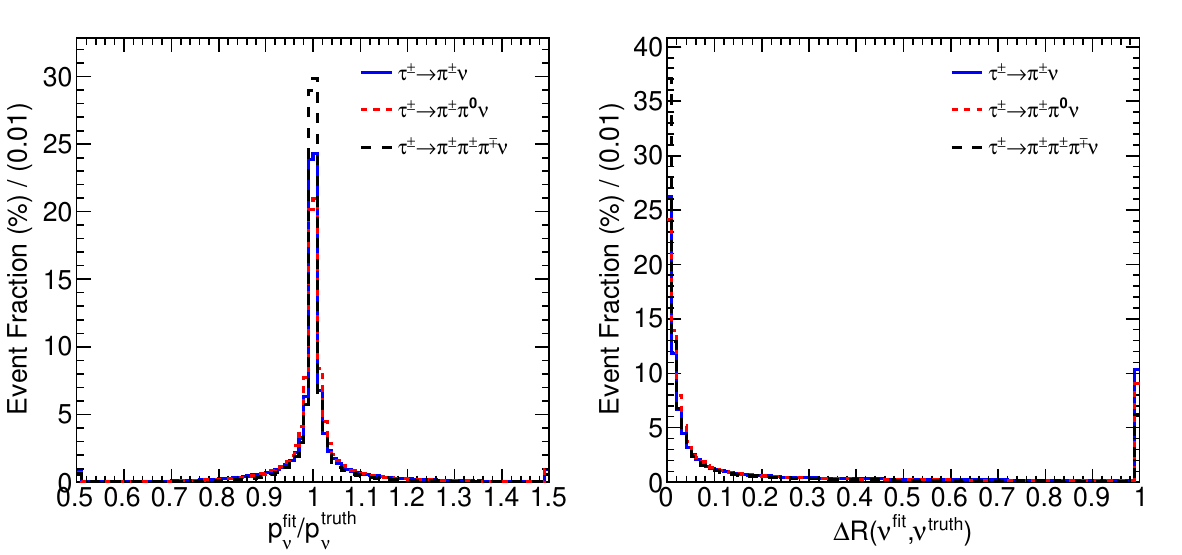}
\caption{The ratio of fitted neutrino momentum to the true one (left), and the $\Delta R$ distance between them (right). The left and/or right most bins contain the overflows.}
\label{fig:reco}
\end{figure}

\begin{figure}[!tb]
\centering
\includegraphics[width=0.45\textwidth]{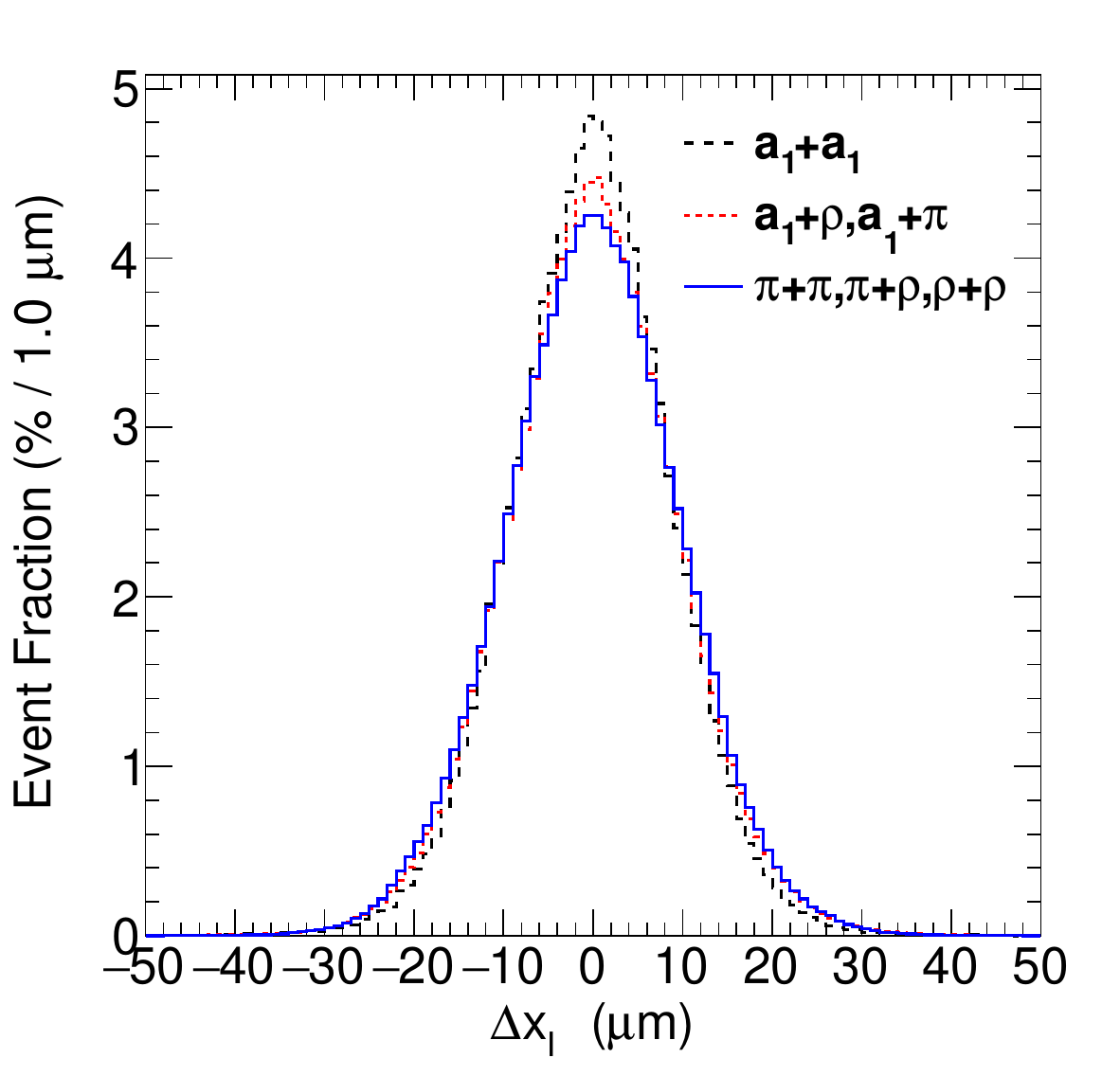}
\includegraphics[width=0.45\textwidth]{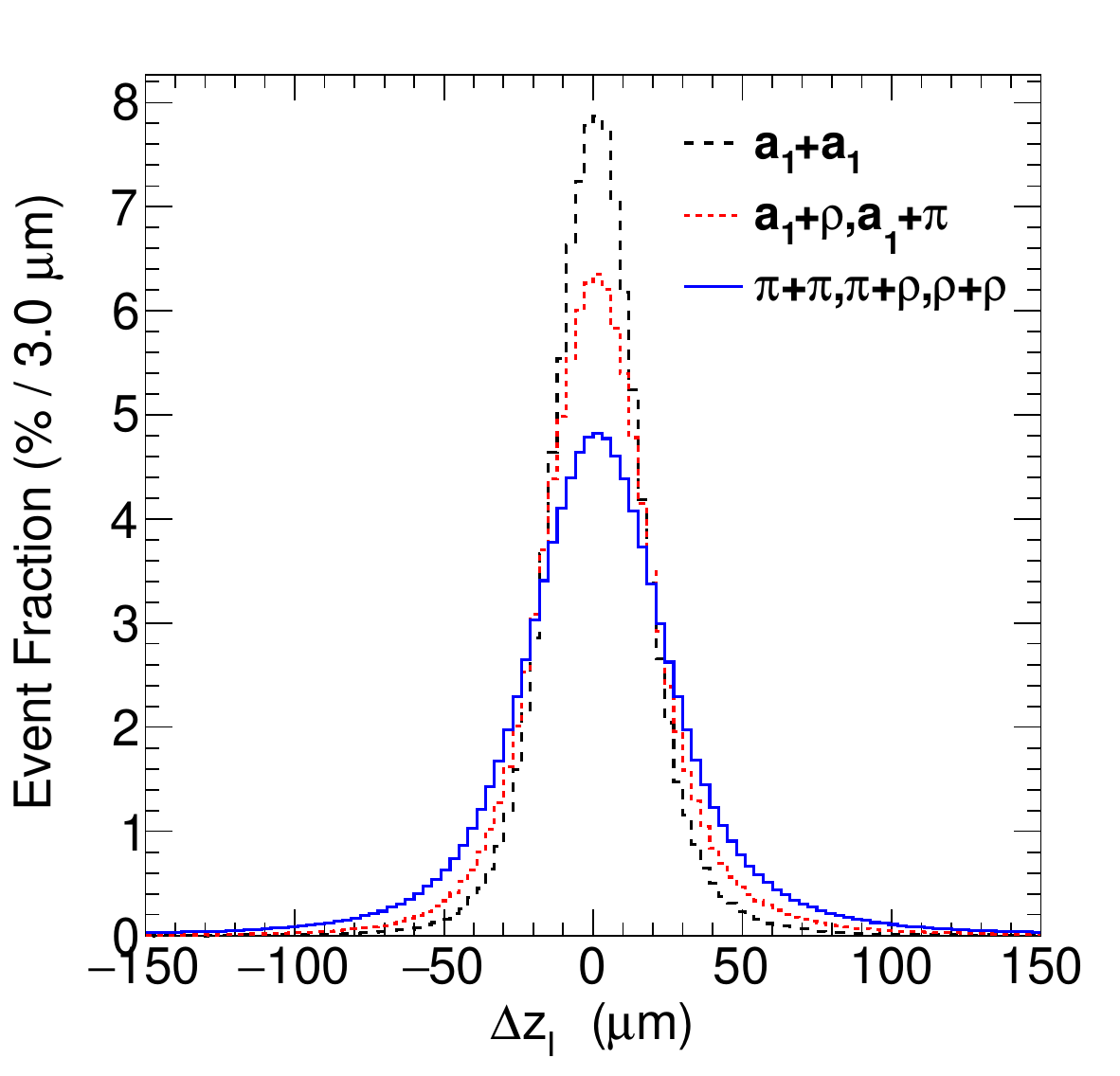}
\caption{The distributions of $\Delta x_{\text{I}}$ ($x_{\text{I}}^{\text{fit}}$- $x_{\text{I}}^{\text{truth}}$) and  $\Delta z_{\text{I}}$ ($z_{\text{I}}^{\text{fit}}$- $z_{\text{I}}^{\text{truth}}$) after the fit in different $\tau^+\tau^-$ decay modes.}
\label{fig:IP}
\end{figure}

In~\autoref{fig:reco}, the ratio of fitted to true neutrino momenta (left panel), and the $\Delta R=\Delta\eta \oplus \Delta\phi$ distance between them (right panel), are shown for different tau decay modes. It is clear that the neutrinos from tau decay can be well reconstructed, and the 3-prong mode (through $a_1$) has the best precision which is considered for the first time in this kind of searches and thus can further improve the search sensitivity in~\cite{tau_exp1,Abdallah:2003xd}.
The resolutions of the IP after the fit are shown in~\autoref{fig:IP}. The post-fit resolutions of $\Delta z_{\text{I}}$ are about 16, 20 and 27 $\mu m$ for six, four and two-track final states, respectively. They indicate some improvements with respect to $\sigma_z$ resolutions in~\autoref{tab:sigma}.


\begin{figure}[!tb]
\centering
\includegraphics[width=0.9\textwidth]{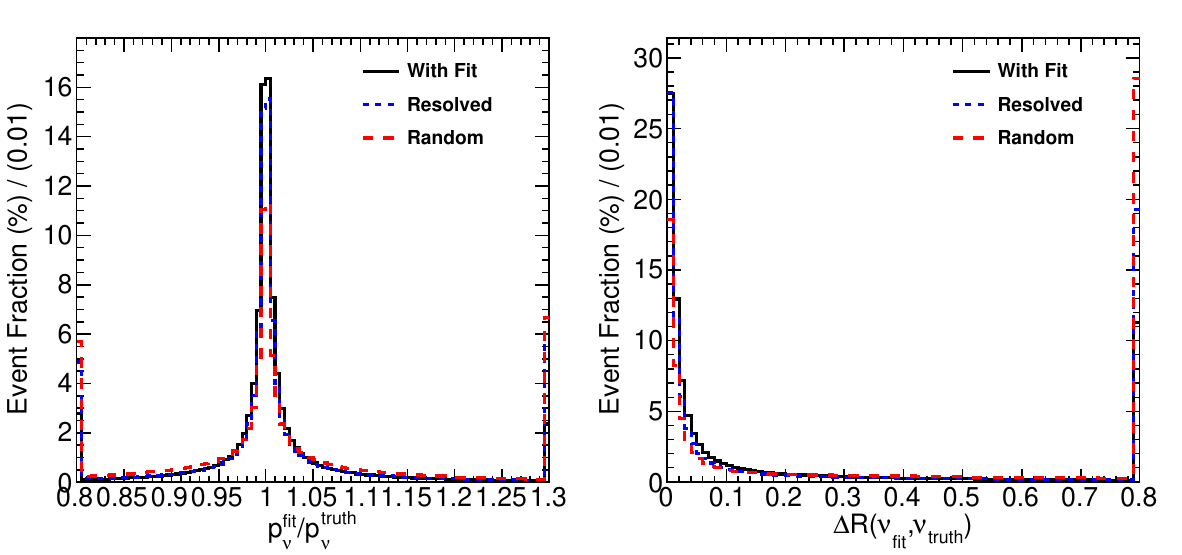}
\caption{The ratio of the fitted neutrino momentum to the true one (left), and the $\Delta R$ distance between them (right), for all ditau decay mode combined. The left and/or right most bins contain the overflows. Three cases are compared: the nominal fit based on \autoref{eq:eq4} (black solid), the randomly selected solution from the two-fold ambiguity (red dashed), and the solution based on the minimum constraint from \autoref{eq:eq4} (blue solid). }
\label{fig:reco_comp2}
\end{figure}

To see the improvement brought by our method, three cases are compared in \autoref{fig:reco_comp2} in terms of the neutrino reconstruction quality. The black solid histograms correspond to our method (With Fit), while the red dashed ones correspond to a procedure with no fit of the system four-momentum and a solution is randomly chosen from the two-fold ambiguity of the neutrinos (Random). The blue dashed histograms are also obtained without a fit, but used \autoref{eq:chi2_2}-\autoref{eq:chi2_3} to resolve the two-fold ambiguity (Resolved). This is similar to the idea proposed in~\cite{Kuhn}, but has four important differences:
\begin{itemize}
\item Our impact parameters are defined with respect to the origin, whereas \cite{Kuhn} used the closest approach between two tracks, which is not the usual definition nowadays.
\item We assumed curved track trajectories, whereas \cite{Kuhn} and the formulas therein assume straight lines.
\item We incorporated the resolutions of impact parameters and the IP point, which is necessary to get a realistic estimation.
\item We varied the system four-momentum in the fit to account for the ISR/FSR effect.
\end{itemize}

If defining the fraction of events with $0.9<p_\nu^\text{fit}/p_\nu^\text{truth}<1.1$ and $\Delta R(\nu^\text{fit},\nu^\text{truth})<0.2$ as the useful signal events for the EDM and $g$-2 measurements, they turn out to be 64.6\%, 58.6\% and 42.0\% for the ``With Fit", ``Resolved" and ``Random" cases, respectively. Therefore, the improvement on the fraction of useful signal events is 6.0\% (16.6\%) for ``With Fit" (``Resolved") over ``Resolved" (``Random").



\section{The measurement of $\tau$ EDM and $g$-$2$ and model interpretations}
\label{sec:measurement_and_model}
For our own convenience, we express the new physics contributions to the EDM and $g$-$2$ as 
\begin{align}
\label{equ:NPEDMg2}
\mathcal{L}_{d_\ell} &\supset -\frac{i}{2}d_\ell^{NP}\bar{\ell}\sigma_{\mu\nu}\gamma_5\ell F^{\mu\nu} = \frac{i}{2}\frac{\sqrt{2}e}{v}\left(\frac{v}{\Lambda}\right)^2c_\ell^{NP}\bar{\ell}\sigma_{\mu\nu}\gamma_5\ell F^{\mu\nu}, \nonumber \\
\mathcal{L}_{a_\ell} &\supset \frac{e}{4m_\ell}a_\ell^{NP}\bar{\ell}\sigma_{\mu\nu}\ell F^{\mu\nu}.
\end{align} 
with $v= 246$ GeV being the electroweak energy scale, and $\Lambda=1$ TeV. $d_\ell^{NP}=d_\ell - d_\ell^{SM}$ and $a_\ell^{NP}=a_\ell-a_\ell^{SM}$. These contributions can be matched to the general vertex~\autoref{equ:GammaLepton} through~\autoref{equ:EDM_g2_Def}. Note that we also introduce a dimensionless parameter $c_\ell^{NP}$ for the EDM Lagrangian.

\subsection{Matrix Element and the Optimal Observable}
\label{sec:measurement}

With fully reconstructed momenta for all final-state particles, the calculation of the matrix element event by event is possible which, according to our parameterization in~\autoref{equ:NPEDMg2}, has the following form~\footnote{Note that the SM predictions are embedded into $M_0^{d,a}$, hence $d^{NP}(c^{NP}) = 0$ and $a^{NP} = 0$ correspond to the SM case. The limits on $|d_\tau^{NP}|$ and $|a_\tau^{NP}|$ reported later can thus be treated as the uncertainties of the measurement of the SM predictions. }
\begin{align}
\label{equ:ME}
|\mathcal{M}|^2_{d_\tau} &\propto M^d_0 - M^d_1\frac{c^{NP}_\tau}{\Lambda} + M^d_2\left(\frac{c^{NP}_\tau}{\Lambda}\right)^2, \nonumber\\
|\mathcal{M}|^2_{a_\tau} &\propto M^a_0 + M^a_1\frac{a^{NP}_\tau}{2m_\tau} + M^a_2\left(\frac{a^{NP}_\tau}{2m_\tau}\right)^2,
\end{align}
In order to preserve the spin correlation, the Spin Projector~\cite{Jadach:1985ac} is used to calculate $M^{(d,a)}_0,\ M^{(d,a)}_1$ and $M^{(d,a)}_2$ which depend on the momenta of all final-state particles. The tau decay couplings are adapted from~\cite{Hagiwara:2012vz} with the following coupling of the currents to the tau fermion line:
\begin{align}
J_{\pm}^\mu(\tau^\pm\to\pi^\pm\nu) &= p_{\pi^\pm}^\mu, \nonumber \\
J_{\pm}^\mu(\tau^\pm\to\pi^\pm\pi^0\nu) &= p_{\pi^\pm}^\mu - p_{\pi^0}^\mu, \nonumber\\
J_{\pm}^\mu(\tau^\pm\to\pi_1^\pm\pi_2^\pm\pi_3^\mp\nu) &= F^{13}(q_1^\mu-q_3^\mu-G^{13}Q^\mu)+(1\leftrightarrow2), 
\end{align}
where $Q^\mu = q_1^\mu + q_2^\mu + q_3^\mu$, $G^{i3}=\frac{Q\cdot(q_i-q_3)}{Q^2}$ and $F^{i3}$ are the form factors for the $a_1$ channel~\cite{Hagiwara:2012vz}.

\begin{figure}[!tb]
\centering
\includegraphics[width=0.48\textwidth]{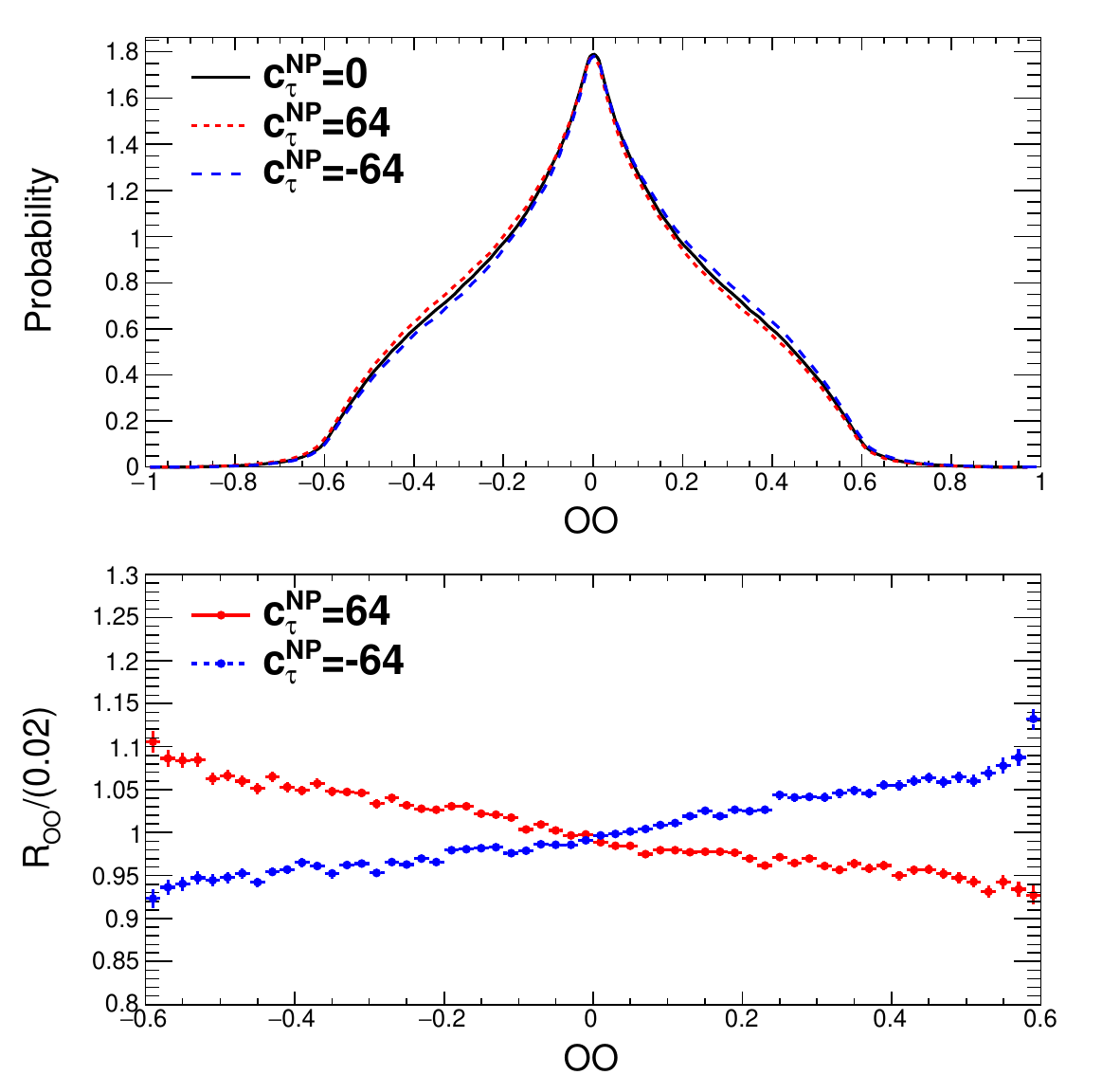}
\includegraphics[width=0.48\textwidth]{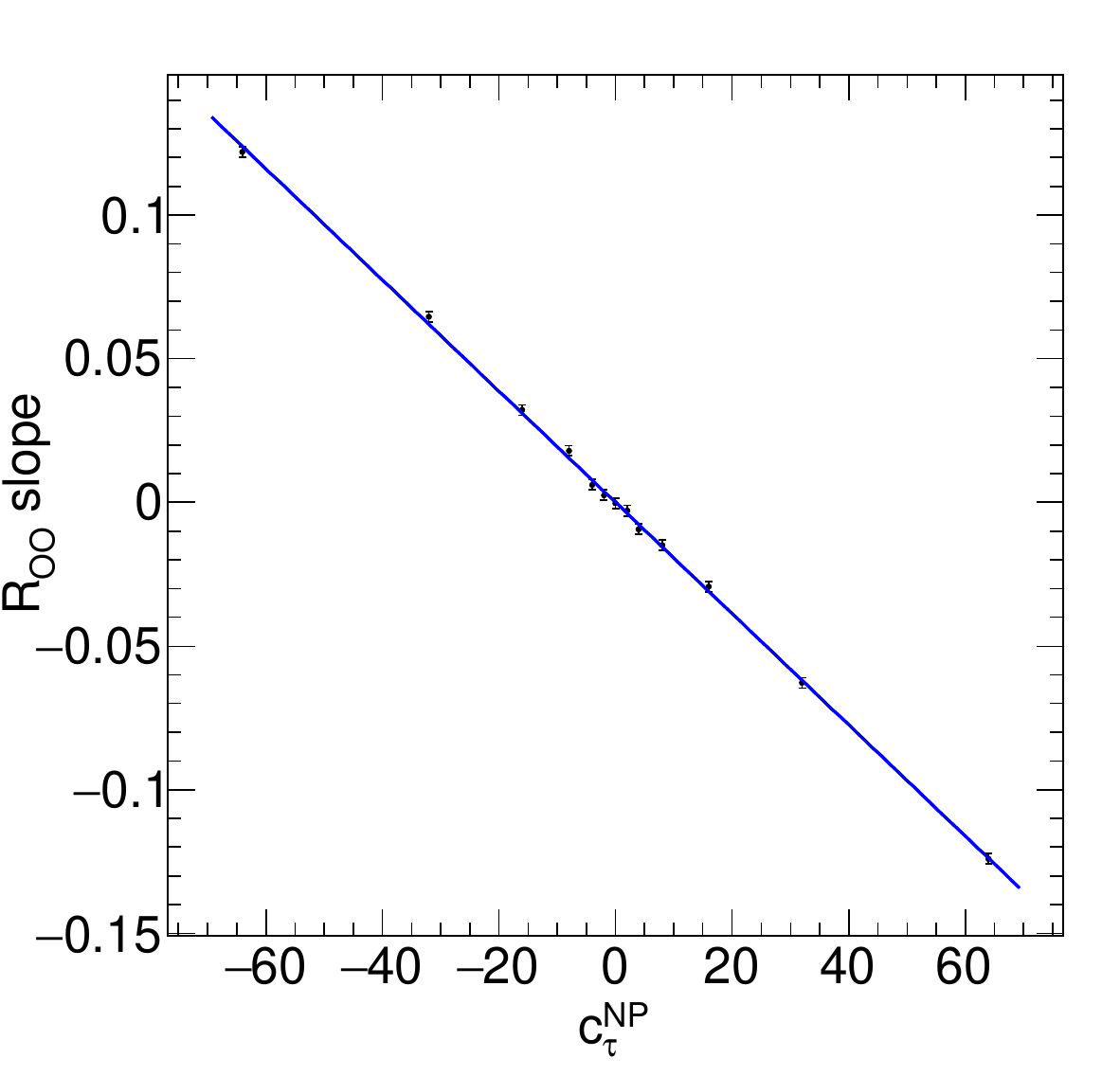}
\caption{The probability distributions of $\mathcal{OO}$ and $R_{\mathcal{OO}}$ as a function of $\mathcal{OO}$ for different $c^{NP}_\tau$'s (left), and the $R_{\mathcal{OO}}$ slope as a function of $c^{NP}_\tau$ (right), where the statistical error bars correspond to 50 $\text{fb}^{-1}$ of data.}
\label{fig:oo_edm}
\end{figure}

\begin{figure}[!tb]
\centering
\includegraphics[width=0.48\textwidth]{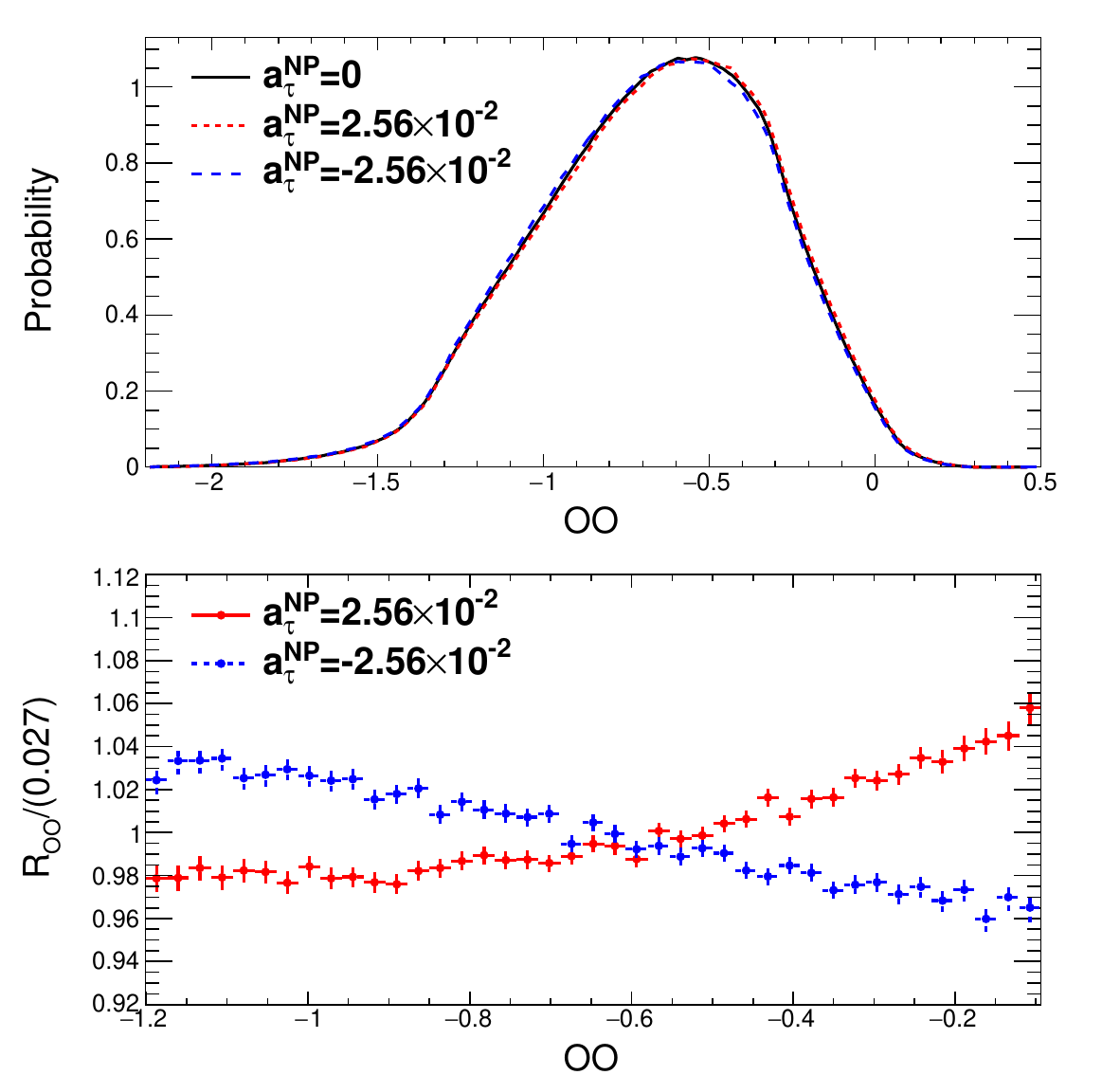}
\includegraphics[width=0.48\textwidth]{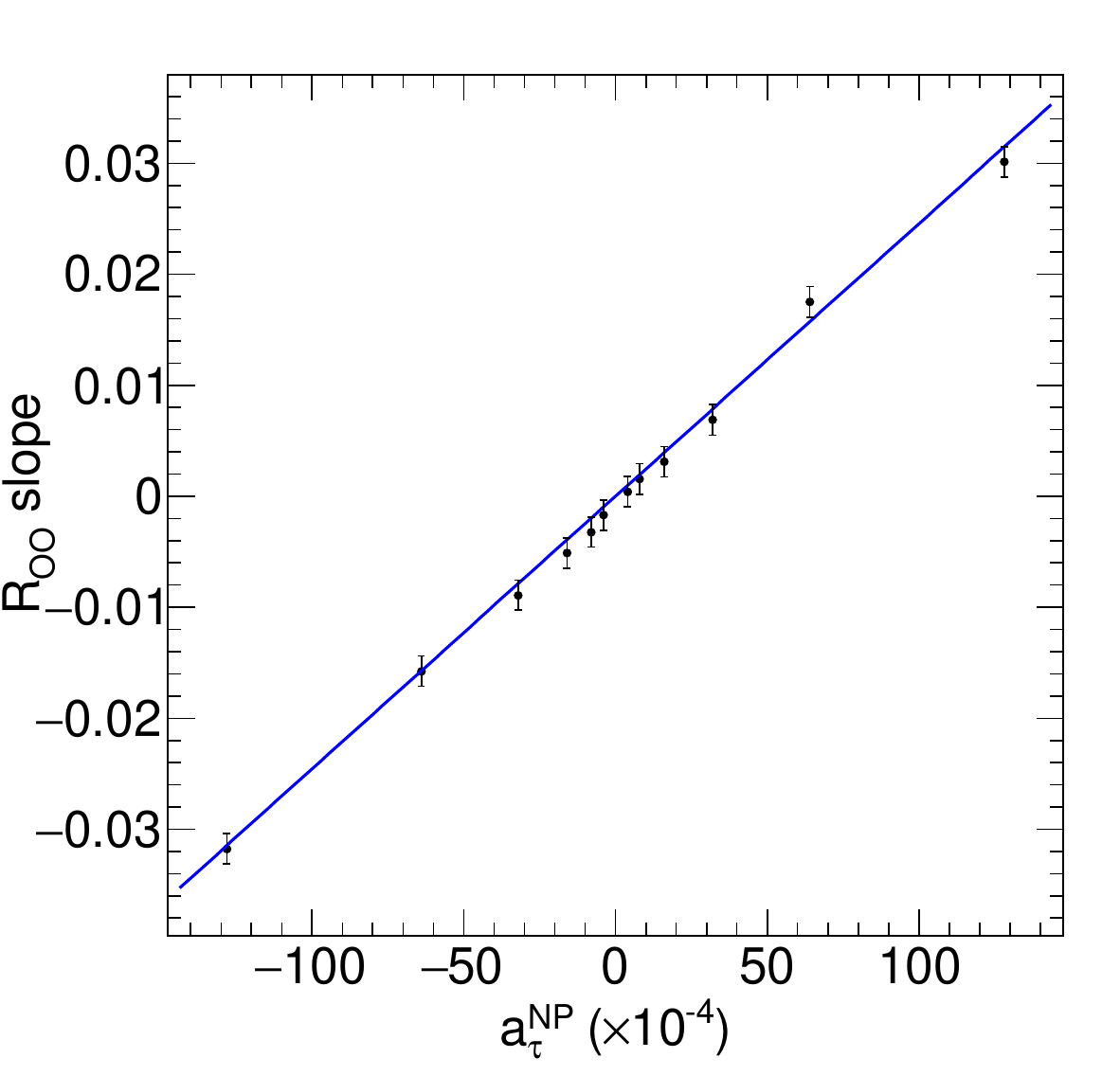}
\caption{The probability distributions of $\mathcal{OO}$ (top) and $R_{\mathcal{OO}}$ as a function of $\mathcal{OO}$ for different $a^{NP}_\tau$'s (left), and the $R_{\mathcal{OO}}$ slope as a function of $a^{NP}_\tau$ (right), where the statistical error bars correspond to 50 $\text{fb}^{-1}$ of data.}
\label{fig:oo_atau}
\end{figure}

Based on the coefficients $M^{(d,a)}_0$ and $M^{(d,a)}_1$, an Optimal Observable ($\mathcal{OO}$)~\cite{Atwood:1991ka,Diehl:1993br} is constructed as 
\begin{equation}
\label{equ:OO}
\mathcal{OO}^{(i)}\equiv \frac{(M^i_1/\text{GeV})}{M^i_0},
\end{equation}
which is sensitive to the values of $c_\tau^{NP}$ and $a_\tau^{NP}$\footnote{Higher-order terms ($M_2^{(d,a)}$) are not considered in this work.}. The $\mathcal{OO}$ distributions for three different choices of $c^{NP}_\tau$ and $a^{NP}_\tau$ are shown in the upper left panels of~\autoref{fig:oo_edm} and \autoref{fig:oo_atau} respectively. Slight shifts in the distributions for different $c^{NP}_\tau$ and $a^{NP}_\tau$ values with respect to $c^{NP}_\tau=0$ and $a^{NP}_\tau=0$ are observed. With the large data statistics that can be cumulated at $Belle$-II and future tau/charm factories, this shift can be detected and provide evidence for the EDM or anomalous $g$-2 of the tau lepton. 

To quantify the shifts, the ratio of $\mathcal{OO}$ distributions between any $c^{NP}_\tau$($a^{NP}_\tau$) hypothesis and 
$c^{NP}_\tau(a^{NP}_\tau)=0$ is first obtained, 
as shown in the lower left panels of \autoref{fig:oo_edm} and \autoref{fig:oo_atau} ($R_\mathcal{OO}$). If there is no BSM contribution, $R_\mathcal{OO}$ would be a flat line at 1. With BSM present, a slope is developed which can be fitted with a linear function of
\begin{equation}
\label{equ:slope}
R_\mathcal{OO} = 1+ b(\mathcal{OO}-x_0),
\end{equation}
where $b$ is the slope, and $x_0$ is the intersection point where the lines cross each other. It is found that $x_0=-0.665$ (0) for $g$-2 (EDM). Once $b$ is fitted for different $a^{NP}_\tau$ (and $c^{NP}_\tau$), their relations can be visualized as in the right panels of \autoref{fig:oo_edm} and \autoref{fig:oo_atau}. Close to $c^{NP}_\tau=0$ and $a^{NP}_\tau=0$, these relations are linear and can be fitted with $b_c = k_c c^{NP}_\tau$ and $b_a = k_a a^{NP}_\tau$. With the fitted 1-$\sigma$ CL error for the slope being $\delta b$, the corresponding 1-$\sigma$ precisions for the BSM parameters are $\delta b_c/k_c$ and $\delta b_a/k_a$ for EDM and $g$-2, respectively. 
The results are listed in \autoref{tab:sensitivity}.

It is worthwhile to note that this analysis is insensitive to absolute event yields, since distributions with alternative $a^{NP}_\tau$ and $c^{NP}_\tau$ are normalized to $a^{NP}_\tau=0$ and $c^{NP}_\tau=0$, respectively, and only shape differences are important when the ratios are taken. Therefore, the results are not expected to be sensitive to higher-order corrections to the $e^+e^-\to\tau^+\tau^-$ signal production cross section.
\begin{table}[!bt]
\centering
\caption{The 1-$\sigma$ CL sensitivities to EDM and $g$-2 parameters for different integrated luminosities. To extrapolate from luminosity $\mathcal{L}_1$ to $\mathcal{L}_2$, a factor of $(\mathcal{L}_1/\mathcal{L}_2)^{1/2}$ is applied.}
\label{tab:sensitivity}
\begin{tabular}{c|ccc}
\hline
 $\mathcal{L}$ & 1 $\text{ab}^{-1}$ & 10 $\text{ab}^{-1}$ & 50 $\text{ab}^{-1}$ \\
\hline
 $|d^{NP}_\tau|$ (e$\cdot$cm) &  $1.44\times 10^{-18}$ & $4.56\times 10^{-19}$ & $2.04\times 10^{-19}$ \\
 $|a^{NP}_\tau|$ & $1.24\times 10^{-4}$ & $3.92\times 10^{-5}$ & $1.75\times 10^{-5}$ \\
\hline
\end{tabular}
\end{table}

To check the extra improvement from the neutrino information, the analysis is repeated following the method in~\cite{tau_exp1,Bernreuther:1993nd}, which uses only the momenta of the visible decay products~\footnote{For simplicity, the two-fold ambiguity in determining the direction of tau pair~\cite{tau_exp1,Bernreuther:1993nd} is ignored, the tau momentum reconstructed using our method is used instead, since following exactly the methods in~\cite{tau_exp1,Bernreuther:1993nd} is beyond our scope. Thus, in this comparison, we are in the conservative side.}, the matrix element is recalculated which has a form similar to that in \autoref{equ:ME} but is averaged over unobserved momenta.
New optimal observables are then calculated for the same events in the $\pi\pi,\ \pi\rho,\ \rho\rho$ modes. It is found that the sensitivity for $d_\ell^{NP}$ is a factor of four better with our method.


\subsection{Constraints on BSM}
\label{sec:model}

\begin{figure}[!hbt]
\centering
\raisebox{4mm}{\includegraphics[width=0.34\textwidth]{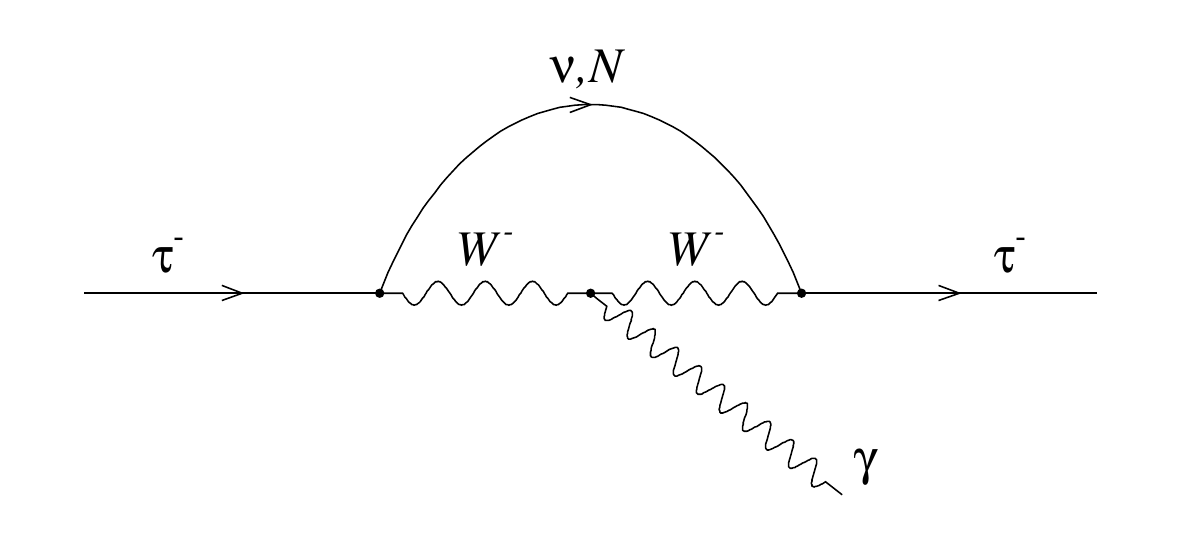}}
\put(-80, 0){\textbf{(a)}} 
\raisebox{4mm}{\includegraphics[width=0.34\textwidth]{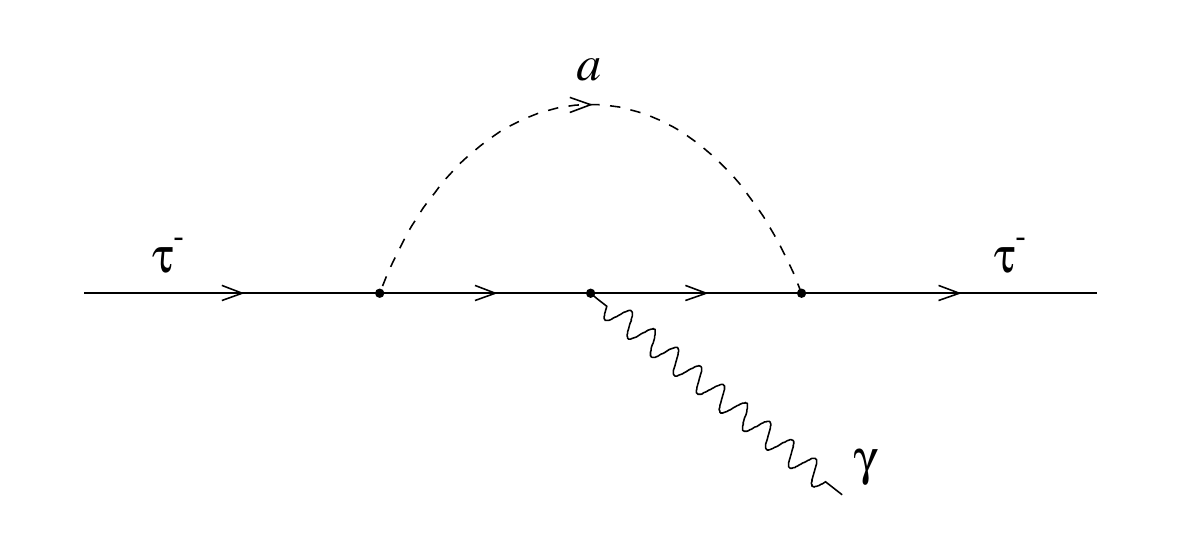}}
\put(-80, 0){\textbf{(b)}}
\includegraphics[width=0.34\textwidth]{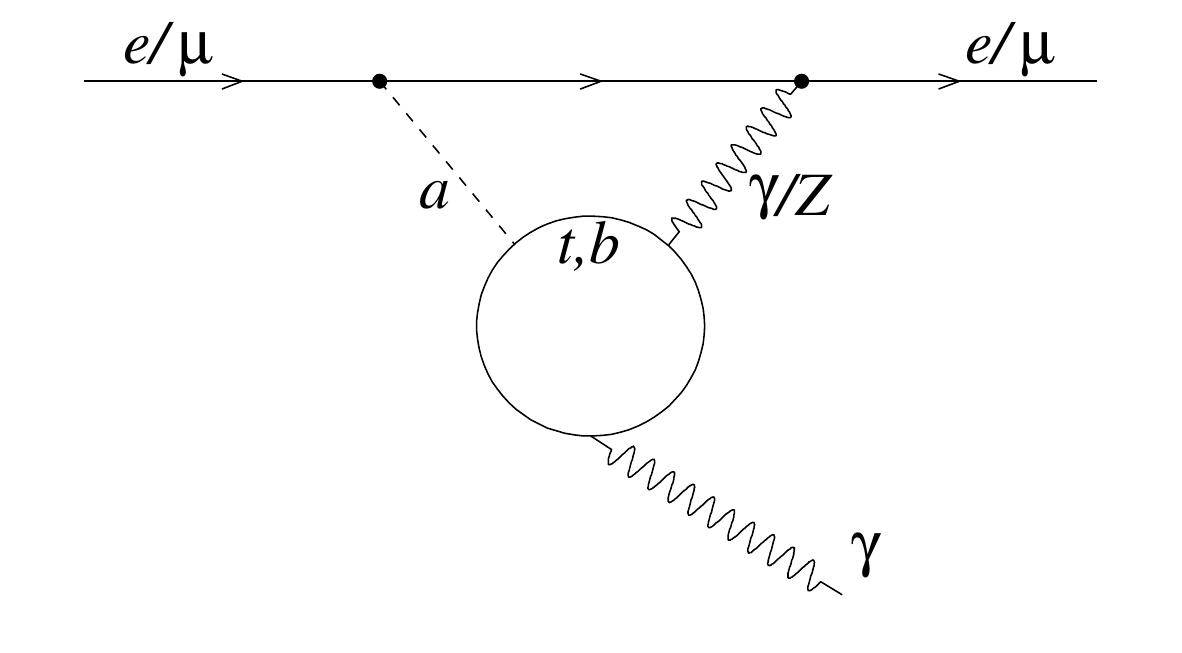}
\put(-80, 0){\textbf{(c)}}
\caption{The Feynman diagram mediated by a heavy mirror neutrino (a), a light scalar (b) and the 2-loop Bar-Zee diagram (c) contributing to the EDM and $g$-2.}
\label{fig:feyn}
\end{figure}

\subsubsection{A heavy mirror neutrino}

\begin{figure}[!hbt]
\centering
\includegraphics[width=0.5\textwidth]{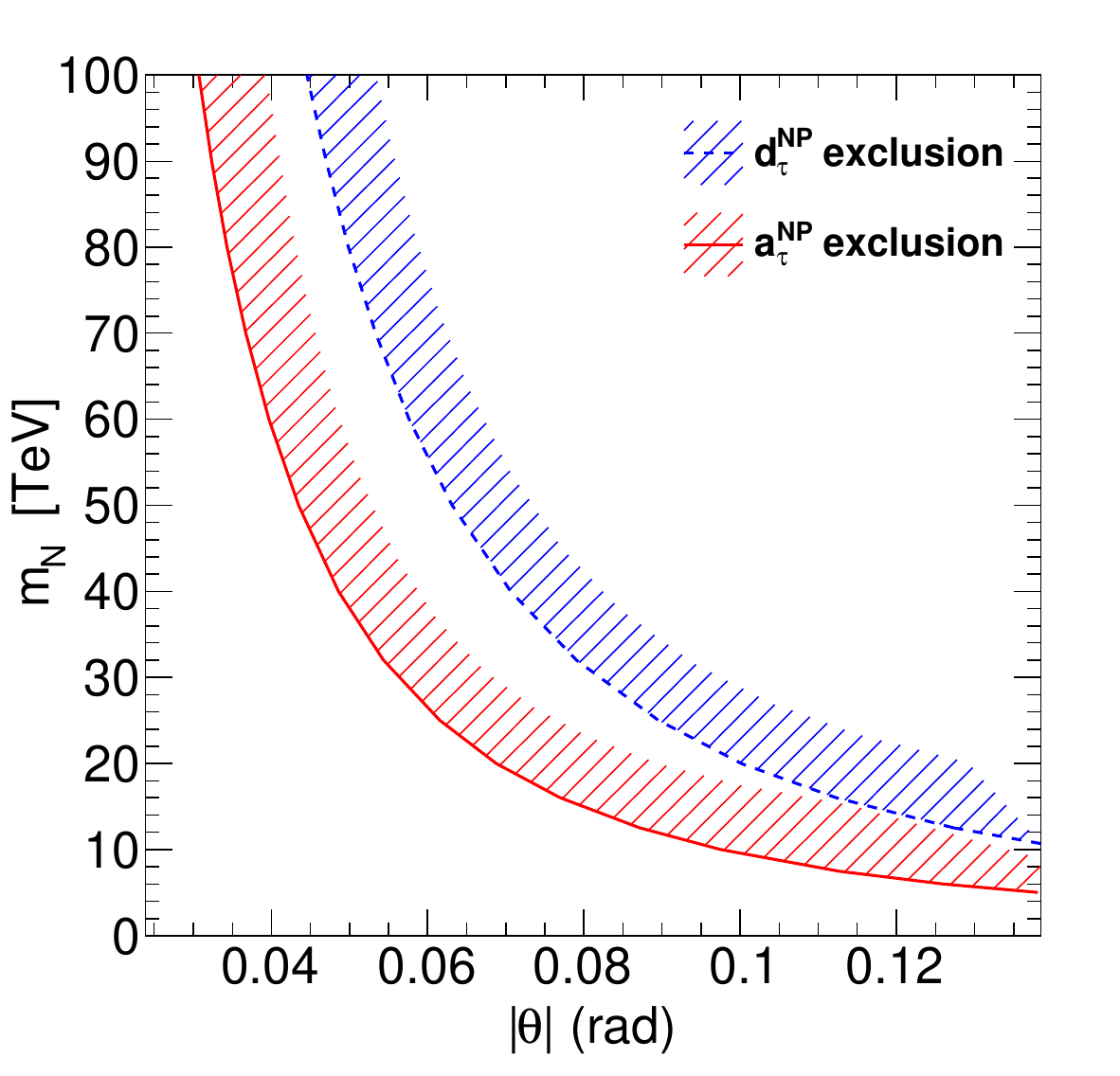} \\
\caption{The 95\% exclusion of a mirror neutrino model in the mixing angle-$m_N$ plane, for both tau EDM ($\Delta\chi=\pi/2$) and $g$-2 ($\Delta\chi=0$). }
\label{fig:sens_mirror}
\end{figure}

Large tau EDM and $g$-2 can be realized in models with mirror leptons arising from GUT, extended SUSY or Kaluza-Klein theories. These particles have $V+A$ type of couplings to the SM leptons, and mixing among them is possible. The Feynman diagram contributing to the tau EDM and $g$-2 is shown in \autoref{fig:feyn}(a). The bi-unitary transformations between the weak and mass eigenstates of the leptons~\cite{Ibrahim:2010va} can be parametrized as
\begin{equation}
 D^{\tau,\nu}_{L,R} =\left(\begin{array}{cc}
\cos\theta & e^{-i\chi}\sin\theta \\
-e^{i\chi}\sin\theta & \cos\theta 
\end{array}\right)\equiv D(\chi),
\end{equation}
where $\theta$ is the mixing angle and $\chi$ is a CP phase. 
The same mixing angle $\theta$ in the two sectors is not necessary, but chosen here for simplicity, i.e.,
\begin{equation}
\begin{array}{cccc}
\left(
\begin{array}{c}
\tau \\
E_\tau 
\end{array}
\right) = D(\chi_1)
\left(
\begin{array}{c}
\tau_1 \\
\tau_2 
\end{array}
\right),
&  &  &
\left(
\begin{array}{c}
\nu \\
N
\end{array}
\right) = D(\chi_2)
\left(
\begin{array}{c}
\nu_1 \\
\nu_2 
\end{array}
\right),
\end{array}
\end{equation}
where $\tau_1$ and $\nu_1$ are the light tau and neutrino mass eigenstates, $\tau_2$ and $\nu_2$ are their heavy mirror eigenstates. With this setup, the charged current with the $W$ boson reads
\begin{eqnarray}
J^\mu & = & (\cos^2\theta+\sin^2\theta e^{-i\chi})\bar{\nu}_1\gamma^\mu\tau_1
                  - (\cos^2\theta-\sin^2\theta e^{-i\chi})\bar{\nu}_1\gamma^\mu \gamma_5\tau_1 \nonumber \\
          &    & + \cos\theta\sin\theta(e^{i\chi}-1)\bar{\nu}_2\gamma^\mu\tau_1 
                    - \cos\theta\sin\theta(e^{i\chi}+1)\bar{\nu}_2\gamma^\mu\gamma_5\tau_1
\end{eqnarray}
The EDM or $g$-2 is most conveniently calculated in the 't Hooft-Feynman gauge~\cite{Fujikawa:1972fe}. Neglecting terms proportional to $m_\tau/m_W$ or $m_\tau/m_N$, the EDM can be calculated as
\begin{equation}
d^{NP}_\tau = \frac{eG_F m_N}{4\sqrt{2}\pi^2}\int_0^1 dz\frac{(1-z)(4-4z+rz)\sin^2\theta\cos^2\theta\sin\Delta\chi}{1-z+rz},
\end{equation}
where $r\equiv m_N^2/m_W^2$, $G_F$ is the Fermi constant and $\Delta\chi = \chi_1-\chi_2$. When $\Delta\chi=\pm\pi/2$ and $m_N\gg m_W$, maximum EDM can be achieved:
\begin{equation}
\left| d^{NP}_\tau \right| = \frac{eG_F m_N \cos^2\theta \sin^2\theta}{8\sqrt{2}\pi^2}.
\end{equation}
The mirror neutrino mass enhancement is because of a factor $m_N/m_W$ in the goldstone boson vertex. This actually enables us to exclude the mirror neutrino mass from above, as shown by the blue curve in \autoref{fig:sens_mirror}. 

As the CP phase difference $\Delta\chi$ goes from $\pi/2$ to 0, $|d^{NP}_\tau|$ goes down, and $a^{NP}_\tau$ goes up. In the case of $\Delta\chi=0$, maximum $a^{NP}_\tau$ can be reached:
\begin{equation}
a^{NP}_\tau = \frac{G_F m_\tau^2 \sin^2(2\theta)}{8\sqrt{2}\pi^2}\int_0^1 dz (1-z) \frac{(4-2z)(1-z)+4\frac{m_N}{m_\tau}(1-z)+\frac{m_N^2}{m_W^2}(1+z)z+\frac{m_N^3}{m_W^2m_\tau}z}{1-z+\frac{m_N^2}{m_W^2}z}.
\end{equation}
In the $m_N\gg m_W$ limit, it can be simplified to
\begin{equation}
a^{NP}_\tau = \frac{G_F m_\tau m_N \cos^2\theta \sin^2\theta}{4\sqrt{2}\pi^2},
\end{equation}
and the exclusion is as shown by the red curve in \autoref{fig:sens_mirror}. For small enough mixing angles that are compatible with observed data, the mirror neutrino mass goes beyond 100 TeV range, which escapes any direct hadron collider search, but can be detected with the tau EDM and/or $g$-2 measurement.

\subsubsection{Light Higgs scalars}
\label{sec:light_scalar}

Enhancement of EDM and $g$-2 could also be observed in a model with extra scalars. One example is the two Higgs doublet model (2HDM) with an extra complex singlet scalar (2HDM+CS) which resembles the scalar sector of the Next to Minimal Supersymmetric SM (NMSSM)~\cite{Keus:2017ioh}. It contains two doublets and one complex singlet, which can be decomposed as
\begin{align}
\Phi_1 = \left(\begin{array}{c}
\phi_1^+\\
\frac{v_1+h_1^0+ia_1^0}{\sqrt{2}}
\end{array}\right), \quad \Phi_2=\left(\begin{array}{c}
\phi_2^+\\
\frac{v_2+h_2^0+ia_2^0}{\sqrt{2}}
\end{array}\right), \quad S = \frac{1}{\sqrt{2}}(\omega + \phi_4 + i\phi_5),
\end{align}
where $v_1$, $v_2$ and $\omega$ are the vacuum expectation value (vev), and $\tan\beta=v_2/v_1$. With $\beta$, one can rotate into the Higgs basis in which only one doublet acquires vev:
\begin{align}
\left(\begin{array}{c}
\hat{\Phi}_1 \\
\hat{\Phi}_2 \\
\hat{S} 
\end{array}\right) = \left(\begin{array}{ccc}
\cos\beta & \sin\beta & 0 \\
-\sin\beta & \cos\beta & 0 \\
0 & 0 & 1 
\end{array}\right)\left(\begin{array}{c}
\Phi_1 \\
\Phi_2 \\
S
\end{array}\right)
\end{align}
where 
\begin{align}
\hat{\Phi}_1 = \left(\begin{array}{c}
G^+ \\
\frac{v+\phi_1+iG^0}{\sqrt{2}}
\end{array}\right),\quad \hat{\Phi}_2 = \left(\begin{array}{c}
H^+ \\
\frac{\phi_2 + i\phi_3}{\sqrt{2}}
\end{array}\right),\quad \hat{S}=\frac{1}{\sqrt{2}}(\omega+\phi_4+i\phi_5)
\end{align}

In general, the neutral scalars will mix by a matrix $R$: $\phi_i = R_{ij}h_j$, it contains ten mixing angles, $\theta_{12-15}$, $\theta_{23-25}$, $\theta_{34,35}$, $\theta_{45}$, among which five induce CP-violation: $\theta_{13,15,23,25,34}$. The scalar-gauge coupling are $y_W^{h_i} = \frac{2m_W^2}{v}R_{1i}$ and $y_Z^{h_i}=\frac{m_Z^2}{v}R_{1i}$. 
The Yukawa sector is similar to that in 2HDM, $y_{d}^{h_i} = \frac{m_d}{v}\left(R_{1i}+\xi_d(R_{2i}+iR_{3i})\right)$, $y_{\ell}^{h_i}=\frac{m_\ell}{v}\left(R_{1i}+\xi_\ell(R_{2i}+iR_{3i})\right)$, $y_{u}^{h_i} = \frac{m_u}{v}\left(R_{1i}+\xi_u(R_{2i}-iR_{3i})\right)$. This work focuses on the Type-II Yukawa coupling, where down-type and charged lepton couple to $\Phi_2$ and up-type couples to $\Phi_1$, and $\xi_{d,\ell} = -\tan\beta$, $\xi_u = \cot\beta$.

As claimed in \cite{Bernabeu:2007rr}, when the BSM scale $\Lambda^2$ is much higher than $q^2$ of the collision process, which is the case in the mirror neutrino model, terms of higher orders of $q^2/\Lambda^2$ can be neglected in the form factor expansion. However, the light scalar mass considered here is of $O(1~\text{GeV})$, which is lower than the $Belle$-II collision energy. In this case, what one measures are not the $d_\ell$ or $a_\ell$ themselves, but rather their form factors at a particular $q^2$, and an imaginary part can also develop. The contribution from the one-loop process (\autoref{fig:feyn}(b)) to the real part of the magnetic form factor ($F_2$ in~\autoref{equ:GammaLepton}) for each scalar $a$ is
%
%
\begin{align}
\Re[F_2(\theta)] = - \frac{|y_\ell|^2}{4\pi^2 \sinh\theta} \int_0^1 dx \int_0^{\frac{\theta}{2}}dy \frac{ x^3-kx^2 }{x^2+z(1-x) \left(\frac{\sinh y}{\sinh \frac{\theta}{2}}\right)^2 },
\label{eq:eq20}
\end{align}
%
where $y_\ell$ is the lepton-scalar coupling constant, $k=2[\Re(y_\ell)]^2/|y_\ell|^2$, $z=m_a^2/m_\tau^2$, $\theta=\ln[(1+\beta)/(1-\beta)]/2$ with $\beta$ being the tau velocity in the ditau center-of-mass frame. 
%
%
It is evident from \autoref{eq:eq20} that as the center-of-mass energy $\sqrt{s}$ increases (or $\beta\to 1$), $\Re(F_2)$ will eventually drop to zero. However, the tau and scalar's mass will slow down the running considerably (as opposed to the case of an electron or muon). For example, at $\theta=1.75486$ corresponding to the $\Upsilon(4S)$ center-of-mass energy, and with $m_a=1$ GeV, $\Re(F_2)$ only drops to about 90\% of its initial value at $\theta=0$, the same value as computed for an on-shell photon as $q^2\to 0$. Apart from BSM, the running of the SM QED form factor itself is also interesting on its own, through its momentum and lepton flavor dependence. In the case of a tau lepton, the QED $\Re(F_2)$ at the $\Upsilon(4S)$ energy will drop to about a quarter of the Schwinger correction value \cite{Bernabeu:2007rr}.

\begin{figure}[!hbt]
\centering
\includegraphics[width=0.45\textwidth]{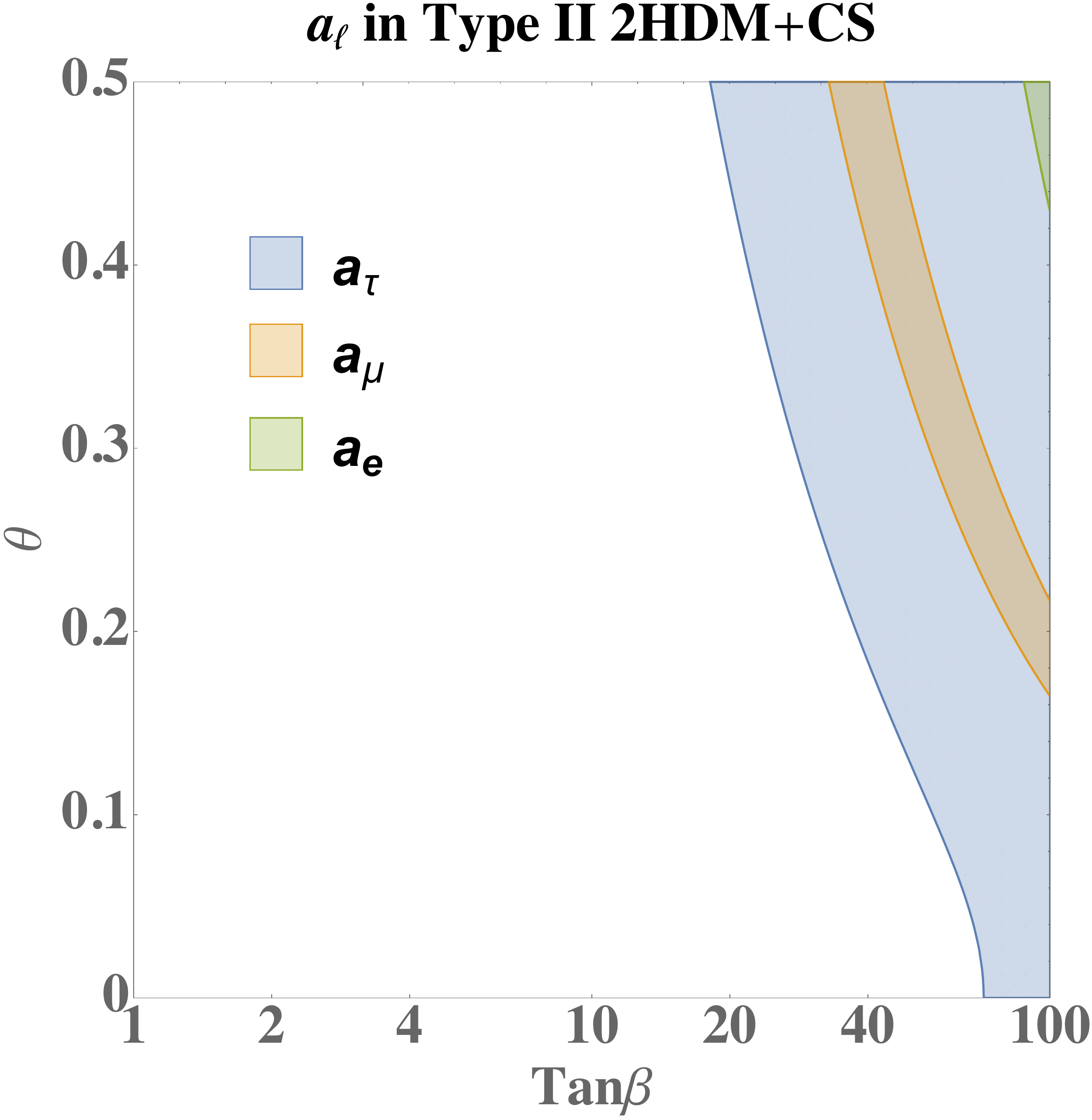}
\put(-160, 40){\textbf{(a)}}
\includegraphics[width=0.45\textwidth]{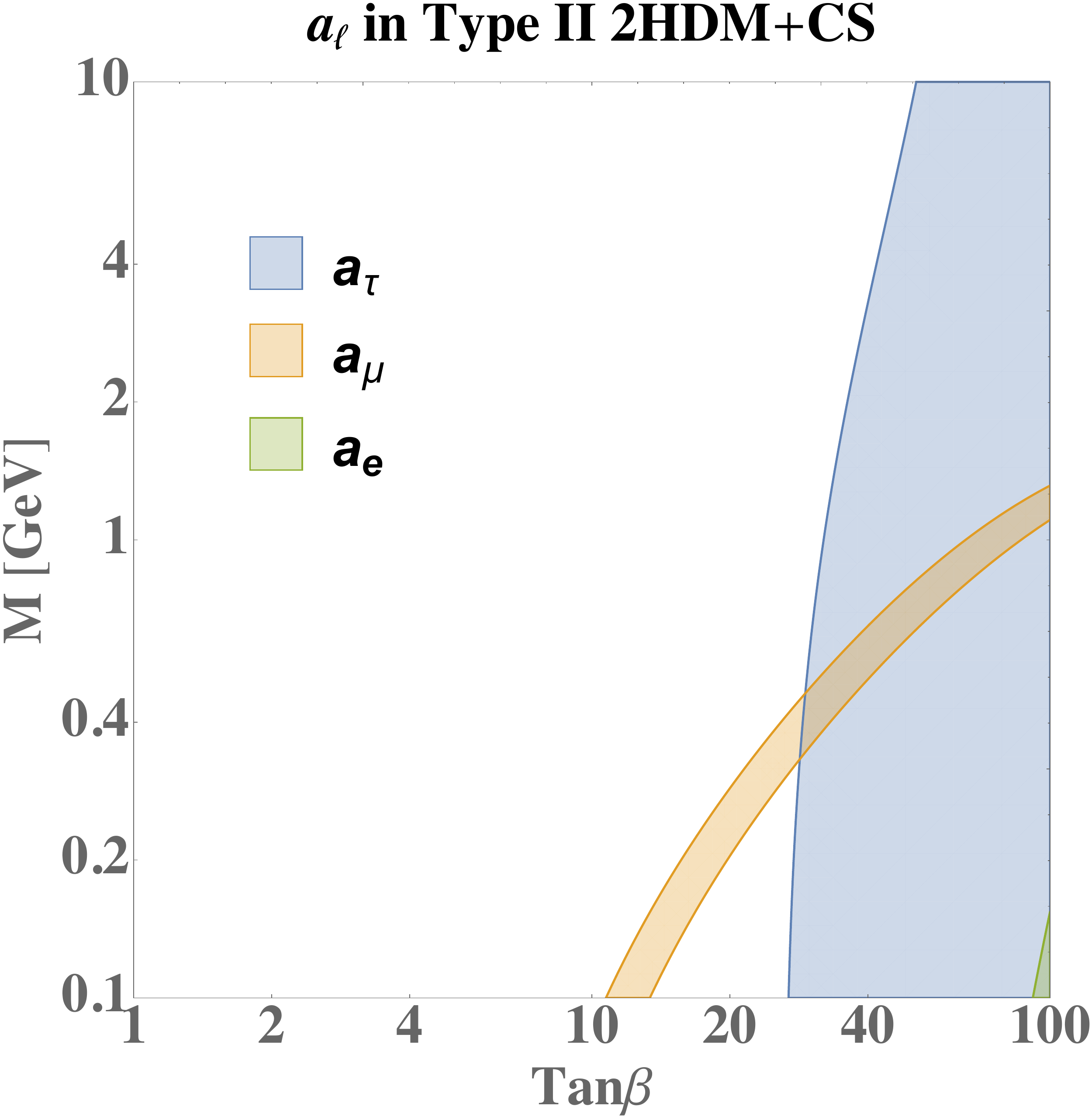}
\put(-160, 40){\textbf{(b)}}\\
\caption{The 95\% exclusion from the $g$-2 measurement in 2HDM+CS model in $\tan\beta$-$\theta$ plane (a) where the extra scalar masses are fixed at 1 GeV, and in $\tan\beta$-$M$ plane (b) where $\theta$ is fixed to be 0.25. The blue and green regions are excluded by the $\tau$ and $e$ measurement respectively. The orange band is the region that could explain the muon $g$-2 anomaly.}
\label{fig:NMSSM}
\end{figure}

When the extra scalars are around GeV scale, their contributions to $g$-2 can be significant. 
Figure~\ref{fig:NMSSM} shows the exclusion region from $g$-2 measurement in $\tan\beta$-$\theta$ plane (left panel) and $\tan\beta$-$M$ plane (right panel). Here, for simplicity, no CP-violation is assumed ($\theta_{13,15,23,25,34}=0$) and $\theta\equiv \theta_{14,24,35}$~\footnote{In our choice of the parameter space, the effects from $\theta_{12,45}$ can be ignored, hence they are also set to 0.}. $M$ is the mass of the extra scalars ($M\equiv m_{h_4} = m_{h_5}$). The mass of other exotic scalars is fixed at 200 GeV, and the main contributions come from the lightest scalars ($h_4$ and $h_5$). In the left panel, the mass of these light scalars is fixed at 1 GeV, and in the right panel, the mixing angle $\theta$ is fixed at 0.25. For the electron and muon, the 2-loop Bar-Zee diagrams as shown in \autoref{fig:feyn}(c) are also important, and the explicit formulas in \cite{ElectronCoupling} are used to calculate their contributions to the magnetic moment.


From \autoref{fig:NMSSM}, it is evident that, in this kind of models, the tau $g$-2 measurement (blue region) is more sensitive than the electron $g$-2 measurement (green region). Further, the improved precision of the tau $g$-2 measurement also covers the region that can explain the muon $g$-2 anomaly (orange band), this puts significant constraints on those models that can be used to explain the muon $g$-2 anomaly.

\section{Summary}
\label{sec:summary}

In conclusion, a new method is proposed in this paper to reconstruct the neutrinos from the tau decays. The fraction of signal events with good fitted neutrinos improves from $42.0\%$ based on pure kinematics, to $58.6\%$ with the impact parameters resolving the two-fold ambiguity in solutions. It is further improved to $64.6\%$ with fits including the impact parameters constraints. With all the final states from tau decay being reconstructed, the matrix element for each event is calculated and employed to obtain the sensitivity on the extra contributions to EDM and $g$-2 of the tau lepton produced in the low energy $e^+e^-$ colliders. Under this framework, it is predicted that with 50 $\text{ab}^{-1}$ of data that will be collected by $Belle$-II, a precision of $|d^{NP}_\tau|<2.04\times 10^{-19}$ e$\cdot$cm can be achieved for the EDM, and $|a^{NP}_\tau|<1.75\times 10^{-5}$ for the $g$-2 of tau (about $1.5\%$ of the SM prediction), when systematics are not considered. 

The improved sensitivities on EDM and $g$-2 are used to constrain two representative models. In the model with mirror leptons, the results can exclude heavy mirror neutrinos with a mass of $O$(100 TeV) that has $V$+$A$ coupling to SM leptons through mixing, which is hard to be directly produced at hadron colliders. The mirror neutrino can cause observable anomalies in both tau EDM and $g$-2 (a simultaneous search for both without overlap is possible).
On the other hand, the $g$-2 measurement with the improved precision can also explore the parameter space in the 2HDM+CS model which resembles the scalar sector of the NMSSM. In this model, it is found that the $g$-2 measurement of tau lepton is more sensitive than the measurement of electron in some parameter spaces, and it can also constrain some parameter space that can explain the muon $g$-2 anomaly.

There are two points worth discussion. 
First, we are not giving precise theoretical predictions of EDM and $g$-2 values, but rather, we are estimating the experimental accuracy that can be achieved. We expect that in the actual analysis, the systematics due to higher order effects \cite{Had} on the shapes of distributions will be evaluated with respect to the accuracy of Monte Carlo generator at that time, which is not the subject of this work.
Second, the tau decay simulation is based on \cite{Hagiwara:2012vz}, which has been tested against TAUOLA \cite{TAUOLA}. The accuracy of the tau decay modeling can be improved with the $Belle$-II data over the time, and the remaining discrepancy with the data constitutes systematics of the search, which is also not covered in this work.

This new framework and the large amount of data collected at current and future tau factories will significantly improve the precision of the tau EDM and $g$-2 measurement, and hence can provide a new opportunity to constrain BSM with much better sensitivity.


\begin{acknowledgments}
X. Chen is supported by the National Natural Science Foundation of China (grant  11675087) and Tsinghua University Initiative Scientific Research Program. Y. Wu is supported by the Natural Sciences and Engineering Research Council of Canada.
\end{acknowledgments}

\bibliographystyle{bibsty}
\bibliography{references}

\providecommand{\href}[2]{#2}\begingroup\raggedright\begin{thebibliography}{10}

\bibitem{Zhao:2014vga}
S.-M. Zhao, T.-F. Feng, X.-J. Zhan, H.-B. Zhang, and B.~Yan, {\it {The study of
  lepton EDM in CP violating BLMSSM}},  {\em JHEP} {\bf 07} (2015) 124,
  [\href{http://arxiv.org/abs/1411.4210}{{\tt arXiv:1411.4210}}].

\bibitem{Yamanaka:2014nba}
N.~Yamanaka, T.~Sato, and T.~Kubota, {\it {Linear programming analysis of the
  $R$-parity violation within EDM-constraints}},  {\em JHEP} {\bf 12} (2014)
  110, [\href{http://arxiv.org/abs/1406.3713}{{\tt arXiv:1406.3713}}].

\bibitem{Appelquist:2004mn}
T.~Appelquist, M.~Piai, and R.~Shrock, {\it {Lepton dipole moments in extended
  technicolor models}},  {\em Phys. Lett.} {\bf B593} (2004) 175--180,
  [\href{http://arxiv.org/abs/hep-ph/0401114}{{\tt hep-ph/0401114}}].

\bibitem{Ibrahim:2008gg}
T.~Ibrahim and P.~Nath, {\it {An MSSM Extension with a Mirror Fourth
  Generation, Neutrino Magnetic Moments and LHC Signatures}},  {\em Phys. Rev.}
  {\bf D78} (2008) 075013, [\href{http://arxiv.org/abs/0806.3880}{{\tt
  arXiv:0806.3880}}].

\bibitem{Ibrahim:2010va}
T.~Ibrahim and P.~Nath, {\it {Large Tau and Tau Neutrino Electric Dipole
  Moments in Models with Vector Like Multiplets}},  {\em Phys. Rev.} {\bf D81}
  (2010) 033007, [\href{http://arxiv.org/abs/1001.0231}{{\tt
  arXiv:1001.0231}}]. [Erratum: Phys. Rev.{\bf D89} (2014) 119902].

\bibitem{Ilakovac:2013wfa}
A.~Ilakovac, A.~Pilaftsis, and L.~Popov, {\it {Lepton Dipole Moments in
  Supersymmetric Low-Scale Seesaw Models}},  {\em Phys. Rev.} {\bf D89} (2014)
  015001, [\href{http://arxiv.org/abs/1308.3633}{{\tt arXiv:1308.3633}}].

\bibitem{Eidelman}
S.~Eidelman, D.~Epifanov, M.~Fael, L.~Mercolli, and M.~Passera, {\it {$\tau$
  dipole moments via radiative leptonic $\tau$ decays}},  {\em JHEP} {\bf 03}
  (2016) 140, [\href{http://arxiv.org/abs/1601.07987}{{\tt arXiv:1601.07987}}].

\bibitem{ACME}
{\bf ACME} Collaboration, J.~Baron et~al., {\it {Order of Magnitude Smaller
  Limit on the Electric Dipole Moment of the Electron}},  {\em Science} {\bf
  343} (2014) 269--272, [\href{http://arxiv.org/abs/1310.7534}{{\tt
  arXiv:1310.7534}}].

\bibitem{Andreev:2018ayy}
{\bf ACME} Collaboration, V.~Andreev et~al., {\it {Improved limit on the
  electric dipole moment of the electron}},  {\em Nature} {\bf 562} (2018)
  355--360.

\bibitem{g_minus_2a}
{\bf Muon g-2} Collaboration, G.~W. Bennett et~al., {\it {Final Report of the
  Muon E821 Anomalous Magnetic Moment Measurement at BNL}},  {\em Phys. Rev.}
  {\bf D73} (2006) 072003, [\href{http://arxiv.org/abs/hep-ex/0602035}{{\tt
  hep-ex/0602035}}].

\bibitem{Miller:2007kk}
J.~P. Miller, E.~de~Rafael, and B.~L. Roberts, {\it {Muon (g-2): Experiment and
  theory}},  {\em Rept. Prog. Phys.} {\bf 70} (2007) 795,
  [\href{http://arxiv.org/abs/hep-ph/0703049}{{\tt hep-ph/0703049}}].

\bibitem{g_minus_2b}
T.~Blum et~al., {\it {The Muon (g-2) Theory Value: Present and Future}},
  \href{http://arxiv.org/abs/1311.2198}{{\tt arXiv:1311.2198}}.

\bibitem{Grange:2015fou}
{\bf Muon g-2} Collaboration, J.~Grange et~al., {\it {Muon (g-2) Technical
  Design Report}},  \href{http://arxiv.org/abs/1501.06858}{{\tt
  arXiv:1501.06858}}.

\bibitem{Chapelain:2017syu}
{\bf Muon g-2} Collaboration, A.~Chapelain, {\it {The Muon g-2 experiment at
  Fermilab}},  {\em EPJ Web Conf.} {\bf 137} (2017) 08001,
  [\href{http://arxiv.org/abs/1701.02807}{{\tt arXiv:1701.02807}}].

\bibitem{Abe:2019thb}
M.~Abe et~al., {\it {A New Approach for Measuring the Muon Anomalous Magnetic
  Moment and Electric Dipole Moment}},  {\em PTEP} {\bf 2019} (2019), no.~5
  053C02, [\href{http://arxiv.org/abs/1901.03047}{{\tt arXiv:1901.03047}}].

\bibitem{tau_exp1}
{\bf Belle} Collaboration, K.~Inami et~al., {\it {Search for the electric
  dipole moment of the tau lepton}},  {\em Phys. Lett.} {\bf B551} (2003)
  16--26, [\href{http://arxiv.org/abs/hep-ex/0210066}{{\tt hep-ex/0210066}}].

\bibitem{Abdallah:2003xd}
{\bf DELPHI} Collaboration, J.~Abdallah et~al., {\it {Study of tau-pair
  production in photon-photon collisions at LEP and limits on the anomalous
  electromagnetic moments of the tau lepton}},  {\em Eur. Phys. J.} {\bf C35}
  (2004) 159--170, [\href{http://arxiv.org/abs/hep-ex/0406010}{{\tt
  hep-ex/0406010}}].

\bibitem{Ananthanarayan:1994af}
B.~Ananthanarayan and S.~D. Rindani, {\it {Measurement of the tau electric
  dipole moment using longitudinal polarization of e+ e- beams}},  {\em Phys.
  Rev.} {\bf D51} (1995) 5996--6007,
  [\href{http://arxiv.org/abs/hep-ph/9411399}{{\tt hep-ph/9411399}}].

\bibitem{Ananthanarayan:2002fh}
B.~Ananthanarayan, S.~D. Rindani, and A.~Stahl, {\it {CP violation in the
  production of tau leptons at TESLA with beam polarization}},  {\em Eur. Phys.
  J.} {\bf C27} (2003) 33--41, [\href{http://arxiv.org/abs/hep-ph/0204233}{{\tt
  hep-ph/0204233}}].

\bibitem{Bernabeu:2004ww}
J.~Bernabeu, G.~A. Gonzalez-Sprinberg, and J.~Vidal, {\it {CP violation and
  electric-dipole-moment at low energy tau-pair production}},  {\em Nucl.
  Phys.} {\bf B701} (2004) 87--102,
  [\href{http://arxiv.org/abs/hep-ph/0404185}{{\tt hep-ph/0404185}}].

\bibitem{Bernabeu:2006wf}
J.~Bernabeu, G.~A. Gonzalez-Sprinberg, and J.~Vidal, {\it {CP violation and
  electric-dipole-moment at low energy tau production with polarized
  electrons}},  {\em Nucl. Phys.} {\bf B763} (2007) 283--292,
  [\href{http://arxiv.org/abs/hep-ph/0610135}{{\tt hep-ph/0610135}}].

\bibitem{Bernabeu:2007rr}
J.~Bernabeu, G.~A. Gonzalez-Sprinberg, J.~Papavassiliou, and J.~Vidal, {\it
  {Tau anomalous magnetic moment form-factor at super B/flavor factories}},
  {\em Nucl. Phys.} {\bf B790} (2008) 160--174,
  [\href{http://arxiv.org/abs/0707.2496}{{\tt arXiv:0707.2496}}].

\bibitem{Bernabeu:2008ii}
J.~Bernabeu, G.~A. Gonzalez-Sprinberg, and J.~Vidal, {\it {Tau spin
  correlations and the anomalous magnetic moment}},  {\em JHEP} {\bf 01} (2009)
  062, [\href{http://arxiv.org/abs/0807.2366}{{\tt arXiv:0807.2366}}].

\bibitem{Atag:2010ja}
S.~Atag and A.~A. Billur, {\it {Possibility of Determining $\tau$ Lepton
  Electromagnetic Moments in ${\gamma\gamma \to \tau^{+}\tau^{-}}$ Process at
  the CERN-LHC}},  {\em JHEP} {\bf 11} (2010) 060,
  [\href{http://arxiv.org/abs/1005.2841}{{\tt arXiv:1005.2841}}].

\bibitem{Billur:2013rva}
A.~A. Billur and M.~Koksal, {\it {Probe of the electromagnetic moments of the
  tau lepton in gamma-gamma collisions at the CLIC}},  {\em Phys. Rev.} {\bf
  D89} (2014) 037301, [\href{http://arxiv.org/abs/1306.5620}{{\tt
  arXiv:1306.5620}}].

\bibitem{Ozguven:2016rst}
Y.~Özgüven, S.~C. İnan, A.~A. Billur, M.~Köksal, and M.~K. Bahar, {\it
  {Search for the anomalous electromagnetic moments of tau lepton through
  electron–photon scattering at CLIC}},  {\em Nucl. Phys.} {\bf B923} (2017)
  475--490, [\href{http://arxiv.org/abs/1609.08348}{{\tt arXiv:1609.08348}}].

\bibitem{Koksal:2017nmy}
M.~Köksal, S.~C. İnan, A.~A. Billur, Y.~Özgüven, and M.~K. Bahar, {\it
  {Analysis of the anomalous electromagnetic moments of the tau lepton in
  $\gamma p$ collisions at the LHC}},  {\em Phys. Lett.} {\bf B783} (2017) 375,
  [\href{http://arxiv.org/abs/1711.02405}{{\tt arXiv:1711.02405}}].

\bibitem{Patrignani:2016xqp}
{\bf Particle Data Group} Collaboration, M.~Tanabashi et~al., {\it {Review of
  Particle Physics}},  {\em Phys. Rev.} {\bf D98} (2018) 030001.

\bibitem{Eidelman:2007sb}
S.~Eidelman and M.~Passera, {\it {Theory of the tau lepton anomalous magnetic
  moment}},  {\em Mod. Phys. Lett.} {\bf A22} (2007) 159--179,
  [\href{http://arxiv.org/abs/hep-ph/0701260}{{\tt hep-ph/0701260}}].

\bibitem{BelleII}
{\bf Belle-II} Collaboration, T.~Abe et~al., {\it {Belle II Technical Design
  Report}},  \href{http://arxiv.org/abs/1011.0352}{{\tt arXiv:1011.0352}}.

\bibitem{Comp-Phys-Com-64-275-1991}
S.~Jadach, J.~H. Kuhn, and Z.~Was, {\it {TAUOLA: A Library of Monte Carlo
  programs to simulate decays of polarized tau leptons}},  {\em Comput. Phys.
  Commun.} {\bf 64} (1990) 275--299.

\bibitem{Tsai:1971vv}
Y.-S. Tsai, {\it {Decay Correlations of Heavy Leptons in
  $e^++e^-\to\ell^++\ell^-$}},  {\em Phys. Rev.} {\bf D4} (1971) 2821.
  [Erratum: Phys. Rev.{\bf D13} (1976) 771].

\bibitem{CEPC_hcp}
X.~Chen and Y.~Wu, {\it {Search for CP violation effects in the $h\to \tau\tau$
  decay with $e^+e^-$ colliders}},  {\em Eur. Phys. J.} {\bf C77} (2017) 697,
  [\href{http://arxiv.org/abs/1703.04855}{{\tt arXiv:1703.04855}}].

\bibitem{Chen:2017nxp}
X.~Chen and Y.~Wu, {\it {Probing the CP-Violation effects in the $h\tau\tau$
  coupling at the LHC}},  {\em Phys. Lett.} {\bf B790} (2019) 332--338,
  [\href{http://arxiv.org/abs/1708.02882}{{\tt arXiv:1708.02882}}].

\bibitem{BelleII_talk}
B.~Golob, ``{The start of the Belle II experiment at the SuperKEKB e+e-
  factory}.''
\newblock {Belle-II Public Talks,
  \url{https://docs.belle2.org/record/1569/files/BELLE2-TALK-CONF-2019-085.pdf}}.

\bibitem{MG5}
J.~Alwall et~al., {\it {The automated computation of tree-level and
  next-to-leading order differential cross sections, and their matching to
  parton shower simulations}},  {\em JHEP} {\bf 07} (2014) 079,
  [\href{http://arxiv.org/abs/1405.0301}{{\tt arXiv:1405.0301}}].

\bibitem{Pythia8}
T.~Sjöstrand, S.~Ask, J.~R. Christiansen, R.~Corke, N.~Desai, P.~Ilten,
  S.~Mrenna, S.~Prestel, C.~O. Rasmussen, and P.~Z. Skands, {\it {An
  Introduction to PYTHIA 8.2}},  {\em Comput. Phys. Commun.} {\bf 191} (2015)
  159--177, [\href{http://arxiv.org/abs/1410.3012}{{\tt arXiv:1410.3012}}].

\bibitem{Hagiwara:2012vz}
K.~Hagiwara, T.~Li, K.~Mawatari, and J.~Nakamura, {\it {TauDecay: a library to
  simulate polarized tau decays via FeynRules and MadGraph5}},  {\em Eur. Phys.
  J.} {\bf C73} (2013) 2489, [\href{http://arxiv.org/abs/1212.6247}{{\tt
  arXiv:1212.6247}}].

\bibitem{EvtGen}
D.~J. Lange, {\it {The EvtGen particle decay simulation package}},  {\em Nucl.
  Instrum. Meth.} {\bf A462} (2001) 152--155.

\bibitem{Delphes}
{\bf DELPHES 3} Collaboration, J.~de~Favereau et~al., {\it {DELPHES 3, A
  modular framework for fast simulation of a generic collider experiment}},
  {\em JHEP} {\bf 02} (2014) 057, [\href{http://arxiv.org/abs/1307.6346}{{\tt
  arXiv:1307.6346}}].

\bibitem{BaBar}
{\bf BABAR} Collaboration, D.~Boutigny et~al., {\it {The BABAR physics book:
  Physics at an asymmetric $B$ factory}},  1998.

\bibitem{Farhi:1977sg}
E.~Farhi, {\it {A QCD Test for Jets}},  {\em Phys. Rev. Lett.} {\bf 39} (1977)
  1587--1588.

\bibitem{Kuhn}
J.~H. Kuhn, {\it {Tau kinematics from impact parameters}},  {\em Phys.Lett.}
  {\bf B313} (1993) 458--460, [\href{http://arxiv.org/abs/hep-ph/9307269}{{\tt
  hep-ph/9307269}}].

\bibitem{MINUIT}
F.~James and M.~Roos, {\it {Minuit: A System for Function Minimization and
  Analysis of the Parameter Errors and Correlations}},  {\em Comput. Phys.
  Commun.} {\bf 10} (1975) 343--367.

\bibitem{Jadach:1985ac}
S.~Jadach and Z.~Was, {\it {QED O($\alpha^3$) Radiative Corrections to the
  Reaction $e^+\ e^-\to\tau^+\ \tau^-$ Including Spin and Mass Effects}},  {\em
  Acta Phys. Polon.} {\bf B15} (1984) 1151. [Erratum: Acta Phys. Polon.{\bf
  B16} (1985) 483].

\bibitem{Atwood:1991ka}
D.~Atwood and A.~Soni, {\it {Analysis for magnetic moment and electric dipole
  moment form-factors of the top quark via $e^+\ e^- \to t\ \bar{t}$}},  {\em
  Phys. Rev.} {\bf D45} (1992) 2405--2413.

\bibitem{Diehl:1993br}
M.~Diehl and O.~Nachtmann, {\it {Optimal observables for the measurement of
  three gauge boson couplings in $e^+\ e^- \to W^+\ W^-$}},  {\em Z. Phys.}
  {\bf C62} (1994) 397--412.

\bibitem{Bernreuther:1993nd}
W.~Bernreuther, O.~Nachtmann, and P.~Overmann, {\it {The CP violating electric
  and weak dipole moments of the tau lepton from threshold to 500-GeV}},  {\em
  Phys. Rev.} {\bf D48} (1993) 78--88.

\bibitem{Fujikawa:1972fe}
K.~Fujikawa, B.~W. Lee, and A.~I. Sanda, {\it {Generalized Renormalizable Gauge
  Formulation of Spontaneously Broken Gauge Theories}},  {\em Phys. Rev.} {\bf
  D6} (1972) 2923--2943.

\bibitem{Keus:2017ioh}
V.~Keus, N.~Koivunen, and K.~Tuominen, {\it {Singlet scalar and 2HDM extensions
  of the Standard Model: CP-violation and constraints from $(g-2)_\mu$ and
  $e$EDM}},  {\em JHEP} {\bf 09} (2018) 059,
  [\href{http://arxiv.org/abs/1712.09613}{{\tt arXiv:1712.09613}}].

\bibitem{ElectronCoupling}
W.~Altmannshofer, J.~Brod, and M.~Schmaltz, {\it {Experimental constraints on
  the coupling of the Higgs boson to electrons}},  {\em JHEP} {\bf 05} (2015)
  125, [\href{http://arxiv.org/abs/1503.04830}{{\tt arXiv:1503.04830}}].

\bibitem{Had}
{\bf Working Group on Radiative Corrections and Monte Carlo Generators for Low
  Energies} Collaboration, S.~Actis et~al., {\it {Quest for precision in
  hadronic cross sections at low energy: Monte Carlo tools vs. experimental
  data}},  {\em Eur. Phys. J.} {\bf C66} (2010) 585--686,
  [\href{http://arxiv.org/abs/0912.0749}{{\tt arXiv:0912.0749}}].

\bibitem{TAUOLA}
O.~Shekhovtsova, T.~Przedzinski, P.~Roig, and Z.~Was, {\it {Resonance chiral
  Lagrangian currents and $\tau$ decay Monte Carlo}},  {\em Phys. Rev.} {\bf
  D86} (2012) 113008, [\href{http://arxiv.org/abs/1203.3955}{{\tt
  arXiv:1203.3955}}].

\end{thebibliography}\endgroup


\end{document}